\documentclass[a4paper,twocolumn,showpacs,superscriptaddress,floatfix,nofootinbib,prd]{revtex4-1}
\usepackage{graphicx}
\usepackage{epsf}
\usepackage{epsfig}
\usepackage{amssymb,amsmath}
\usepackage[usenames]{color}
\usepackage{amssymb}
\usepackage{times}
\usepackage{hyperref}
\hypersetup{
  colorlinks=true,        
  linkcolor=blue,         
  citecolor=cyan,         
}

\newcommand{\cf}{cf.,~}
\newcommand{\ie}{i.e.,~}
\newcommand{\eg}{e.g.,~}
\newcommand{\ms}{\,{\rm ms}}
\newcommand{\Msun}{M_{\odot}}
\newcommand{\mn}[1]{\texttt{#1}}

\begin{document}

\title{Gravitational-wave signal from binary neutron stars: a systematic
  analysis of the spectral properties}
\author{Luciano~Rezzolla}
\affiliation{Institute for Theoretical Physics
  Max-von-Laue-Strasse 1, 60438 Frankfurt, Germany}
\affiliation{Frankfurt Institute for Advanced Studies, Ruth-Moufang-Str. 1,
60438 Frankfurt, Germany}
\author{Kentaro~Takami}
\affiliation{Kobe City College of Technology, 651-2194 Kobe, Japan}
\affiliation{Institute for Theoretical Physics
  Max-von-Laue-Strasse 1, 60438 Frankfurt, Germany}

\begin{abstract}
A number of works have shown that important information on the equation
of state of matter at nuclear density can be extracted from the
gravitational waves emitted by merging neutron-star binaries. We present
a comprehensive analysis of the gravitational-wave signal emitted during
the inspiral, merger and post-merger of 56 neutron-star binaries. This
sample of binaries, arguably the largest studied to date with realistic
equations of state, spans across six different nuclear-physics equations
of state and ten masses, allowing us to sharpen a number of results
recently obtained on the spectral properties of the gravitational-wave
signal. Overall we find that: (i) for binaries with masses differing no
more than $20\%$, the frequency at gravitational-wave amplitude's maximum
is related quasi-universally with the tidal deformability of the two
stars; (ii) the spectral properties vary during the post-merger phase,
with a transient phase lasting a few millisecond after the merger and
followed by a quasi-stationary phase; (iii) when distinguishing the
spectral peaks between these two phases, a number of ambiguities in the
identification of the peaks disappear, leaving a simple and robust
picture; (iv) using properly identified frequencies, quasi-universal
relations are found between the spectral features and the properties of
the neutron stars; (v) for the most salient peaks analytic fitting
functions can be obtained in terms of the stellar tidal deformability or
compactness. Altogether, these results support the idea that the equation
of state of nuclear matter can be constrained tightly when a signal in
gravitational waves from binary neutron stars is detected.
\end{abstract}


\maketitle


\section{Introduction}
\label{sec:intro}

The recent measurement from the advanced interferometric LIGO detectors
\citep{Harry2010} of the first direct gravitational-wave (GW) signal from
what has been interpreted as the inspiral, merger and ringdown of a
binary system of black holes \cite{Abbott2016a} marks, in many respects,
the beginning of GW astronomy. As additional advanced detectors such as
Virgo \citep{Accadia2011_etal}, and KAGRA \citep{Aso:2013} are going to
become operational in the next few years, we are likely to soon witness
also signals from the inspiral and post-merger of neutron-star binaries
or neutron-star--black-hole binaries. Such systems are not only excellent
sources of GWs, but also the most attractive scenario to explain the
phenomenology associated with short gamma-ray bursts (SGRBs). Starting
from the first suggestions that merging neutron stars could be behind
these phenomena \cite{Narayan92,Eichler89} and supported by the
circumstantial evidence coming from numerous astronomical observations
(see \cite{Berger2013b} for a recent review), numerical simulations have
sharpened the contours of this scenario \cite{Shibata99d, Baiotti08,
  Anderson2007, Liu:2008xy, Bernuzzi2011}. What we now know rather
reliably is that the merger of a binary neutron-star (BNS) system
inevitably leads to the formation of a massive metastable object, which
can either collapse promptly to a black hole or survive up to thousands
of seconds \cite{Ravi2014}, emitting gravitational and electromagnetic
radiation \cite{Zhang2001, Metzger:2011, Rezzolla2014b,
  Ciolfi2014}. Furthermore, if the neutron stars have large magnetic
fields and extended magnetospheres, the inspiral can be accompanied by a
precursor electromagnetic signal \cite{Palenzuela2013a}, while the merger
can lead to instabilities \cite{Siegel2013, Kiuchi2015a} and to the
formation of magnetically confined jet structures once a torus is formed
around the black hole \cite{Rezzolla:2011, Paschalidis2014}. Hence, the
prospects of a multimessenger gravitational and electromagnetic signal
are particularly good in the case of merger of binary neutron stars.

In addition to a strong electromagnetic signal, the merger of a
neutron-star binary also promises a GW signal that will contain important
signatures of the equation of state (EOS) of matter at nuclear
densities. These signatures are contained both in the inspiral and in the
post-merger signals. The former is reasonably well understood
analytically \cite{Flanagan08, Baiotti:2010, Bernuzzi2012, Bernuzzi2015,
  Hinderer2016} and can be tracked accurately with advanced high-order
numerical codes \cite{Radice2013b, Radice2013c, Radice2015} and over many
orbits now \cite{Hotokezaka2016}. More importantly, the instantaneous GW
frequency at amplitude maximum $f_{\rm max}$ has been shown to correlate
closely with the tidal deformability of the two stars \cite{Read2013,
  Bernuzzi2014, Takami2015}. The latter part of the signal has already
been studied in the past \cite{Shibata05d, Oechslin07b, Baiotti08}, but
it has become the focus of attention particularly over the last few years
\cite{Bauswein2011, Stergioulas2011b, Bauswein2012, Hotokezaka2013c,
  Bauswein2014, Takami2014, Takami2015, Bernuzzi2015a, Palenzuela2015,
  Bauswein2015, Dietrich2015, Foucart2015, Lehner2016}. This large bulk
of work has reached some generally agreed-upon conclusions, but also has
raised points that are a matter of debate that we hope to clarify here.

Let us therefore start by summarising the aspects of the post-merger GW
signal that seem to be robust and confirmed by several groups employing a
variety of numerical methods and mathematical approximations. Given a
realistic BNS system, namely, with a total mass between $\sim 2.4$ and
$\sim 2.8\,M_{\odot} $, and a mass difference between the two components
that is $\sim 20\%$ or less, then the spectrum of the post-merger GW
signal will present at least three strong peaks \cite{Bauswein2011,
  Stergioulas2011b, Bauswein2012, Hotokezaka2013c, Bauswein2014,
  Takami2014, Takami2015}. These peaks were dubbed $f_1, f_{2}$ and $f_3$
in Refs. \cite{Takami2014,Takami2015} and were found to satisfy the
following approximate relation: $f_{2} \simeq (f_1 + f_3)/2$.  A simple
mechanical toy model was also presented in \cite{Takami2015}, that
provided an intuitive explanation on the origin of these peaks and why
they should be almost equally spaced. In addition to these three peaks,
another peak can be identified in the power spectral density (PSD) of the
GW signal, although not always. This is given by the coupling between the
$\ell=2=m$ fundamental mode (which yields the $f_2$ peak) and a
quasi-radial (fundamental) axisymmetric mode, \ie with $\ell=2,m=0$; this
mode was dubbed $f_{2\mbox{-}0}$ in
Ref. \cite{Stergioulas2011b}. Finally, Ref. \cite{Bauswein2015}
introduced the concept of the $f_{\rm spiral}$ peak frequency and
associated it to a ``rotating pattern of a deformation of spiral
shape''. It is difficult to measure this motion in a numerical-relativity
calculation without the possible contamination of spatial-gauge effects;
however, as we will show in the following, $f_{\rm spiral}$ coincides
with $f_1$ in the large majority of the cases so that, in the end, the
strongest and most robust features of the post-merger signal are confined
to four frequencies: $f_1, f_{2}, f_3$, and $f_{2\mbox{-}0}$,
respectively.

In addition to selecting the most salient spectral features of the
post-merger GW signal, the analyses of Refs. \cite{Bauswein2011,
  Stergioulas2011b, Bauswein2012, Hotokezaka2013c, Bauswein2014,
  Takami2014, Takami2015, Bernuzzi2015a} have also tried to associate the
values of these frequencies to the stellar properties of the two stars
before the merger and hence to their EOS. This was initiated by Refs.
\cite{Bauswein2011, Bauswein2012}, who showed that $f_2$ correlated with
the radius of the maximum-mass nonrotating configuration; the correlation
found was rather tight, but restricted to binaries having all the same
total mass of $2.7\,\Msun$ \cite{Bauswein2011}; later on, the correlation
was explored also when considering a larger sample of masses
\cite{Bauswein2012}, and it was noted that the correlation was not as
tight as previously expected. Indeed, Refs. \cite{Takami2014, Takami2015}
discovered that the correlations of the $f_2$ peaks with the radius of
the maximum-mass nonrotating configuration depend on the total mass of
the system and hence are not ``universal'' in the sense of being only
weakly dependent on the EOS (see also
Refs. \cite{Bauswein2012,Hotokezaka2013c}). At the same time,
Refs. \cite{Takami2014, Takami2015} showed that the $f_1$ peak is
correlated with the total compactness of the stars in a quasi-universal
manner, and highlighted a number of other correlations (24 different ones
were presented in Fig. 15 of \cite{Takami2015}); some of these
correlations had been already presented in the literature, \eg in
Refs. \cite{Read2013, Bernuzzi2014}, while most of them were presented
there for the first time. More specifically, correlations were found
between $f_{\rm max}, f_1$ and $f_2$ frequencies and the physical
quantities of the binary system, \eg the stellar compactness, the average
density, or the dimensionless tidal deformability (some of these
correlations will be further discussed below).

Some of the results reviewed above are robust, while other are less
so. In particular, although the interpretation of the largest peak in the
spectrum is normally attributed to the $\ell=2=m$ mode of the HMNS, the
interpretation of the low-frequency peak (or peaks) is still subject to
debate. More specifically, it is unclear whether such a peak (or peaks)
correlates in a ``universal'' manner with the stellar properties, as
shown in \cite{Takami2015} for the $f_1$ peak, or not, as shown in
\cite{Bauswein2015} for the $f_{\rm spiral}$ peak.

The purpose of this paper is to try and clarify this debate and, more
specifically, to show that for the cases considered here the $f_{\rm
  spiral}$ frequency either coincides with the $f_1$ frequency or falls
in a part of the PSD of the signal where no significant power can be
found, thus explaining why no universal behaviour was found in
Ref. \cite{Bauswein2015}. We reach this conclusion by extending the
sample of binaries considered in Refs. \cite{Takami2014, Takami2015} to
include the fully general-relativistic simulations of the very low-mass
binaries with total mass $2\times 1.2\,M_{\odot}$ and that were suggested
by \cite{Bauswein2015} to be missing in our sample. In addition, we also
consider the largest masses that can be supported for a timescale after
the merger sufficiently long to yield an accurate spectrum; depending on
the EOS, these masses can be as large as $2\times 1.5\,M_{\odot}$. The
large majority of binaries have equal masses, but we consider also four
different instances of unequal-mass binaries with mass difference of
about $20\%$.

This complete sample of binaries, which counts a total of 56 binaries and
doubles the sample presented in \cite{Takami2015}, is arguably the
largest studied to date with nuclear-physics EOSs and in full general
relativity (a sample of comparable size but not in full general
relativity was already presented in Ref. \cite{Bauswein2012}). After a
systematic analysis of the complete sample, it was possible to sharpen a
number of arguments on the spectral properties of the GW signal and that
can be summarised as follows:
\begin{itemize}

\item the GW frequency at amplitude maximum $f_{\rm max}$ is related to
  quasi-universally with the tidal deformability of the two stars; this
  correlation is strong for equal-mass binaries and holds as long as the
  binaries have masses that do not differ of more than $20\%$.

\item the post-merger signal is characterised by a transient phase
  lasting a few millisecond after the merger, which is then followed by
  quasi-stationary phase.

\item the spectral properties of the GW signal vary during the
  post-merger phase with a marked difference between the transient and
  the quasi-stationary phase.

\item spectrograms are particularly useful when selecting spectral
  features in the transient phase because peaks that appear in the short
  transient, may be subdominant when analysed in terms of the full PSD.

\item when distinguishing the spectral peaks between the transient and
  quasi-stationary phases, a number of ambiguities in the identification
  of the peaks disappear, leaving a rather simple and robust picture;

\item the strongest and most robust features of the post-merger signal
  are confined to four frequencies: $f_1, f_{2}, f_3$, and
  $f_{2\mbox{-}0}$, where $f_{2} \simeq (f_1 + f_3)/2$ and
  $f_{2\mbox{-}0}$ is the result of a mode coupling.

\item a number of ``universal'' relations can be found between the main
  spectral features and the physical properties of the neutron stars.

\item for all of the correlations found, simple analytic expressions can
  be given either in terms of the dimensionless tidal deformability or of
  the stellar compactness.
\end{itemize}
When considered as a whole, these results support the idea that the
equation of state of nuclear matter can be tightly constrained when a
strong post-merger signal in GWs is measured.

The paper is organised as follows. Section \ref{sec:ns} provides a very
brief summary of the mathematical and numerical methods used to obtain
our results in full general relativity, while Sect.  \ref{sec:results} is
dedicated to the illustration of our results. In particular, in Sects.
\ref{sec:wp_ts}, \ref{sec:wp_qss}, and \ref{sec:wp_aotfpsd} we
concentrate on the waveform properties coming from transient signals,
quasi-stationary signals and from the analysis of the full PSDs,
respectively. The analysis of the correlations of the spectral signatures
with the stellar properties is instead presented in Sect.
\ref{sec:wp_cwsp}, while Sect. \ref{sec:conclusions} contains our
conclusions and future prospects. Three different appendices offer
details on the full set of binaries considered (Appendix
\ref{appendix_a}), on the analysis of some modes (Appendix
\ref{appendix_b}) or on a two-dimensional fit employed in our analysis
(Appendix \ref{appendix_c}).

\section{Mathematical and numerical Setup}
\label{sec:ns}

The mathematical and numerical setup used for the simulations reported
here is the same discussed in \cite{Takami2014,Takami2015} and presented
in greater detail in other papers \cite{Baiotti08, Baiotti:2009gk,
  Baiotti:2010ka}. For completeness we review here only the basic
aspects, referring the interested reader to the papers above for
additional information. All of our simulations have been performed in
full general relativity using a fourth-order finite-differencing code
\texttt{McLachlan} \cite{Brown:2008sb, Loffler:2011ay}, which solves a
conformal traceless formulation of the Einstein
equations \cite{Nakamura87, Shibata95, Baumgarte99}, with a ``$1+\log$''
slicing condition and a ``Gamma-driver'' shift
condition \citep{Alcubierre02a,Pollney:2007ss}. At the same time, the
general-relativistic hydrodynamics equations are solved using the
finite-volume code \texttt{Whisky} \citep{Baiotti04}, which has been
extensively tested in simulations involving the inspiral and merger of
BNSs \cite{Baiotti08, Baiotti:2009gk, Rezzolla:2010, Baiotti:2010}.

The hydrodynamics equations are solved employing the Harten-Lax-van
Leer-Einfeldt (HLLE) \cite{Harten83} approximate Riemann solver
\cite{Harten83}, which is less accurate but more robust, in conjunction
with a Piecewise Parabolic Method (PPM) for the reconstruction of the
evolved variables \cite{Colella84}. For the time integration of the
coupled set of the hydrodynamic and Einstein equations we have used the
Method of Lines (MOL) in conjunction with an explicit fourth-order
Runge-Kutta method \cite{Rezzolla_book:2013}. In all our simulations we
prescribe a Courant-Friedrichs-Lewy (CFL) factor of $0.35$ to compute the
size of the timestep.

\subsubsection{Grid structure and extents}

We employ an adaptive-mesh refinement (AMR) approach that follows closely
the one adopted in \cite{Baiotti08, Kastaun2013} and where the grid
hierarchy is handled by the \texttt{Carpet} mesh-refinement driver
\cite{Schnetter-etal-03b}. It implements vertex-centered mesh refinement,
also known as the box-in-box method, and allows for regridding during the
calculation as well as multiple grid centres. The timestep on each grid
is set by the Courant condition and by the spatial grid resolution for
that level. Boundary data for finer grids are calculated with spatial
prolongation operators employing fifth-order polynomials and with
prolongation in time employing second-order polynomials.

During the inspiral, a grid with the finest refinement and fully covering
each star is centred at the position of the maximum rest-mass
density. The grid hierarchy is composed of six refinement levels and a
$2:1$ refinement factor for successive levels. The grid resolution varies
from $\ensuremath{\Delta h_5 = 0.15\,\Msun}$ (\ie $\simeq 221\,{\rm m}$)
for the finest level, to $\ensuremath{\Delta h_0 = 4.8\,\Msun}$ (\ie
$\simeq 7.1\,{\rm km}$) for the coarsest level, whose outer boundary is
at $514\,\Msun$ (\ie $\simeq 759\,{\rm km}$). To reduce computational
costs in the case of equal-mass binaries, which represent the large
majority of our sample, the grid structure is then replicated employing a
$\pi$-symmetric, \ie a symmetry of $180$ degrees around the
\ensuremath{z}-axis\footnote{In the case of equal-mass binaries we have
  carried out comparative simulations of the same initial data with and
  without the $\pi$-symmetry being imposed, finding no appreciable
  differences in the position (in frequency) of the main peaks of the
  PSDs. At the same time, the amplitude of such peaks can vary (up to
  $\sim 20\%$) and the PSD without $\pi$-symmetry naturally shows more
  power at higher frequencies as these are not suppressed by the
  symmetry.}. Independently of the mass ratio, the whole grid is set up
to be symmetric with respect to the \ensuremath{(x,y)} plane both for
equal- and unequal-mass binaries, with a reflection symmetry across the
$z=0$ plane, again to reduce computational costs.

The number of grid points across the linear dimension of a star is of the
order of $100$, and this is roughly doubled when the merger has taken
place and a HMNS has been formed. The boundary conditions are chosen to
be ``radiative'' for the metric in order to prevent GWs (or other
numerical perturbations) from scattering back into the grid, and
``static'' for the hydrodynamical variables.

Finally, we recall that in Ref. \cite{Takami2015} we have carried out a
resolution study to assess the influence of the resolution on the
spectral properties. By comparing the results with higher and lower
resolutions, we have found that the use of a ``medium'' resolution of
$\ensuremath{\Delta h_5 = 0.15\,\Msun}$ is sufficient to provide
numerically robust measurements of the peaks, that is, peaks which differ
of a few per-cent only from those obtained with higher resolutions (see
Section IV A of \cite{Takami2015}).

\subsubsection{Equations of state}
\label{sec:EOS}

As in our previous work \cite{Takami2014,Takami2015}, we model the stars
with five ``cold'' (\ie at zero temperature) nuclear-physics EOSs: \ie
APR4 \cite{Akmal1998a}, ALF2 \cite{Alford2005}, SLy \cite{Douchin01}, H4
\cite{GlendenningMoszkowski91} and GNH3 \cite{Glendenning1985}. All of
these EOSs satisfy the current observational constraint on the observed
maximum mass in neutron stars, \ie $2.01\pm0.04\,M_\odot$ obtained for
the pulsar PSR J0348+0432 \cite{Antoniadis2013}. In addition, to validate
the results also across ``hot'' EOSs, we consider two binaries described
by the Lattimer-Swesty EOS \cite{Lattimer91}, with nuclear
compressibility parameter $K = 220\ \mathrm{MeV}$ (LS220); these binaries
were first studied in \cite{Radice2016}, where additional information on
their dynamics can be found.

The nuclear-physics EOSs are normally provided in tabular form, but it is
more convenient numerically to express them in terms of a number of
piecewise polytropes \cite{Read:2009a,Rezzolla_book:2013}. Four different
``pieces'' are normally sufficient to reproduce to good precision most of
the EOSs, with three of the pieces describing the high-density core and
one the crustal region; we refer to Table I of Ref. \cite{Takami2015} for
a list of the properties of the various piecewise polytropes used here.

The cold nuclear-physics EOSs also need to be supplemented by a ``hot''
contribution that accounts for the considerable increase in the internal
energy at the merger. This is normally done through a so-called 
``hybrid EOS'' \cite{Rezzolla_book:2013}, in which an ideal-fluid
component that accounts for the shock heating is added to the cold part
\cite{Janka93}. In practice the total pressure and specific internal
energy are expressed as
\begin{align}
\label{EOS:full_a}
 p &= p_\mathrm{c} + p_\mathrm{th}\,,\\
\label{EOS:full_b}
\epsilon &= \epsilon_\mathrm{c} + \epsilon_\mathrm{th}\,,
\end{align}
where $p_\mathrm{c}, \epsilon_\mathrm{c}$ are given by cold
nuclear-physics EOSs (expressed as piecewise polytropes), while the
``thermal'' part is given by
\begin{align}
\label{EOS:hot}
 p_\mathrm{th} &= \rho\, \epsilon_\mathrm{th}
\left(\Gamma_\mathrm{th} -1 \right)\,,\\
\epsilon_\mathrm{th} & = \epsilon - \epsilon_\mathrm{c} \,,
\end{align}
where $\epsilon$ is obtained through the solution of the hydrodynamics
equations and $\Gamma_\mathrm{th}$ is arbitrary, but constrained
mathematically to be $1 \leq \Gamma_\mathrm{th} \leq 2$. After some
experimentation carried out in \cite{Takami2015}, we have concluded that
values $\Gamma_\mathrm{th}=1.8-2.0$ do not introduce a significant
variance in the spectral properties of the GW signal and we have
therefore chosen $\Gamma_\mathrm{th}=2.0$ for consistency with the single
polytrope (see discussion in Sect. IV B. in \cite{Takami2015}).

\subsubsection{Initial data}

The initial data in our simulations represents quasi-equilibrium
irrotational BNSs and is computed with the multi-domain spectral-method
code \texttt{LORENE} \citep{Gourgoulhon-etal-2000:2ns-initial-data} under
the assumption of a conformally flat spacetime metric. All binaries have
an initial coordinate separation between the stellar centres of
$45\,\mathrm{km}$, which yields at least four orbits (or more) before the
merger.

The choice of the masses for the binaries is constrained by two
considerations. The first one is that, given the substantial
computational costs of these simulations we need to consider masses that
are realistic, rather than masses that give, for instance, the largest
mass difference; in practice, this implies that our masses are around
$1.30\,M_{\odot}$, which indeed we take as our fiducial mass. The second
consideration is that in order to model the post-merger reliably we need
an HMNS that survives for a sufficiently large amount of time before
collapsing to a black hole (\eg at least $t_\mathrm{HMNS} \geq
5000\,\Msun \approx 25\ms$); stated differently, binaries with these EOSs
and masses larger than $\sim 2.70\,M_{\odot}$, are not optimal for the
post-merger analysis although we use them to have the largest possible
sample.

As a result of these considerations, for each cold EOS, we have
considered ten equal-mass binaries with average (gravitational) mass at
infinite separation in the range $\bar{M} \equiv (M_{_A}+M_{_B})/2 =
(1.200 - 1.500)\Msun$ for the APR4, ALF2, GNH3, H4, and SLy
EOSs\footnote{Note that because the highest-mass binaries collapse
  promptly to a black hole, the only spectral information we can use in
  these cases is the one relative to the inspiral, \ie $f_{\rm
    max}$.}. Furthermore, as a complement to our set of equal-mass
binaries, we have also considered four unequal-mass binaries with
$\bar{M}=1.300\,\Msun$ and mass ratio $q \simeq 0.93$ and
$\bar{M}=1.275\,\Msun$ and mass ratio $q \simeq 0.82$; these unequal-mass
binaries have been chosen for the GNH3 and SLy EOSs as examples of stiff
and soft EOSs, respectively. The sample is completed by two equal-mass
binaries with masses $\bar{M}=1.338, 1.372\,\Msun$ and described by the
the hot LS220 EOS. Detailed information on all the models and their
properties is collected in Table \ref{tab:models} of Appendix
\ref{appendix_a}.

\begin{figure*}
\begin{center}
\includegraphics[width=16.0cm,height=18cm]{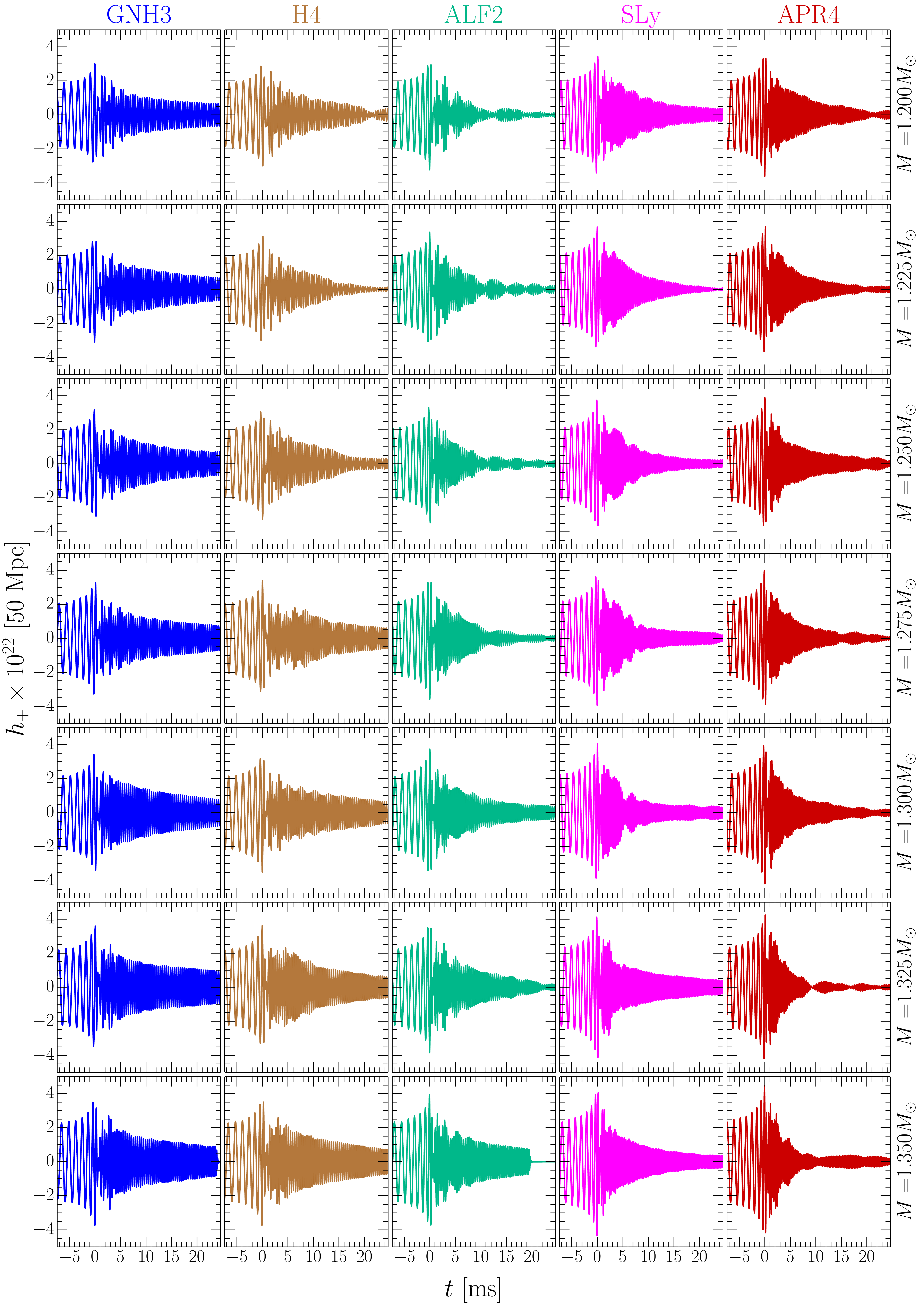}
\caption{Gravitational waveforms for some of the cold-EOS binaries listed
  in Table.~\ref{tab:models}. The rows correspond to the gravitational
  masses, $\bar{M}=1.200, 1.225, 1.250, 1.275, 1.300, 1.325,
  1350\,M_{\odot}$, respectively, while each column refers to a given
  EOS. The different EOSs are distinguished by different colours and we
  will adopt this colour-coding also for all the subsequent plots.  All
  models have formed long-lived HMNSs with $t > 20\,\mathrm{ms}$.}
\label{Waveform_5x7} 
\end{center}
\end{figure*}

\subsubsection{Gravitational-wave signal}
\label{sec:aotgws}

The GW signal is extracted at different surfaces of constant coordinate
radius using the Newman-Penrose formalism, so that the GW polarization
amplitudes $h_+$ and $h_\times$ are related to Weyl curvature scalar
$\psi_4$ by (see Sect.~IV of Ref. \cite{Baiotti08} for details)
\begin{equation}
\ddot{h}_+ - {\rm i} \ddot{h}_\times =
\psi_4 = \sum_{\ell=2}^{\infty}\sum_{m=-\ell}^{\ell} 
\psi_4^{\ell m}\;_{-2}Y_{\ell m}(\theta,\varphi),
\end{equation} 
where the overdot indicates a time derivative and we have introduced the
(multipolar) expansion of $\psi_4$ in spin-weighted spherical
harmonics \cite{Goldberg:1967} of spin-weight $s=-2$. In practice, all of
our analysis  is limited to the dominant mode, \ie the $\ell=m=2$ mode
\begin{equation}
h_{+,\times} 
= \sum_{\ell=2}^{\infty}\sum_{m=-\ell}^{\ell}
 h^{\ell m}_{+,\times} \,{}_{-2}Y_{\ell m}\left(\theta,\varphi\right)
\approx h^{22}_{+,\times} \,{}_{-2}Y_{22}\left(\theta,\varphi\right)\,,
\end{equation}
where ${}_{s}Y_{\ell m}\left(\theta,\varphi\right)$ are the spin-weighted
spherical harmonics. Following previous work \cite{Read2013, Takami2015},
we align the waveforms at the ``time of merger'', which we set to be
$t=0$ and define to be correspondent to the time when the GW amplitude
\begin{equation}
|h| \equiv ( {h^2_+} + {h^2_{\times}} )^{1/2} \,,
\end{equation}
reaches its first maximum. As a result, for most binaries we consider GW
signals in the time interval $t \in [-1500,5000]\, \Msun \approx
[-7.39,24.63] \ms$.  After defining the instantaneous frequency of the GW
as $f_{_\mathrm{GW}} \equiv \frac{1}{2\pi} (d\phi/dt)$, where $\phi =
\arctan( h_{\times} / h_{+} )$ is the phase of the complex gravitational
waveform \cite{Read2013}. The time of the merger is also used to define
the ``frequency at amplitude maximum'' (or peak frequency in
Ref. \cite{Read2013}) as $f_\mathrm{max} \equiv f_{_\mathrm{GW}}(t=0)$.

Another quantity used extensively in our analysis is the PSD of the
effective amplitude and defined as
\begin{eqnarray}
&& \tilde{h}(f) \equiv \sqrt{ \frac{ |\tilde{h}_{+}(f)|^2 +
      |\tilde{h}_{\times}(f)|^2 }{2} } \,,
\end{eqnarray}
with
\begin{eqnarray}
&& \tilde{h}_{+,\times}(f) \equiv 
\left\{
\begin{array}{ll}
\displaystyle
\int 
h_{+,\times}(t)\,e^{-i 2\pi f t} dt & ( f \ge 0 )\\
\displaystyle
0 & ( f < 0 )
\end{array}
\right.\,,
\end{eqnarray}
and where the $+$ and $\times$ indices refer to the two polarization
modes. Using this PSD, we can compute the signal-to-noise ratio (SNR) as
\begin{equation}
\mathrm{SNR} \equiv \left[~ \int ^\infty _0 
\frac{\bigl| 2~\tilde{h}(f) f^{1/2}\bigr|^2}{S_{h}(f)}
~\frac{df}{f}~\right]^{1/2}\,,
\end{equation}
with $S_{h}(f)$ being the noise PSD of a given GW detector [\eg Advanced
  LIGO \cite{url:adLIGO_Sh_curve}, or the Einstein Telescope (ET)
  \cite{Punturo:2010,Punturo2010b}].

\begin{figure*}
\begin{center}
\includegraphics[width=7.5cm,height=7.5cm]
                {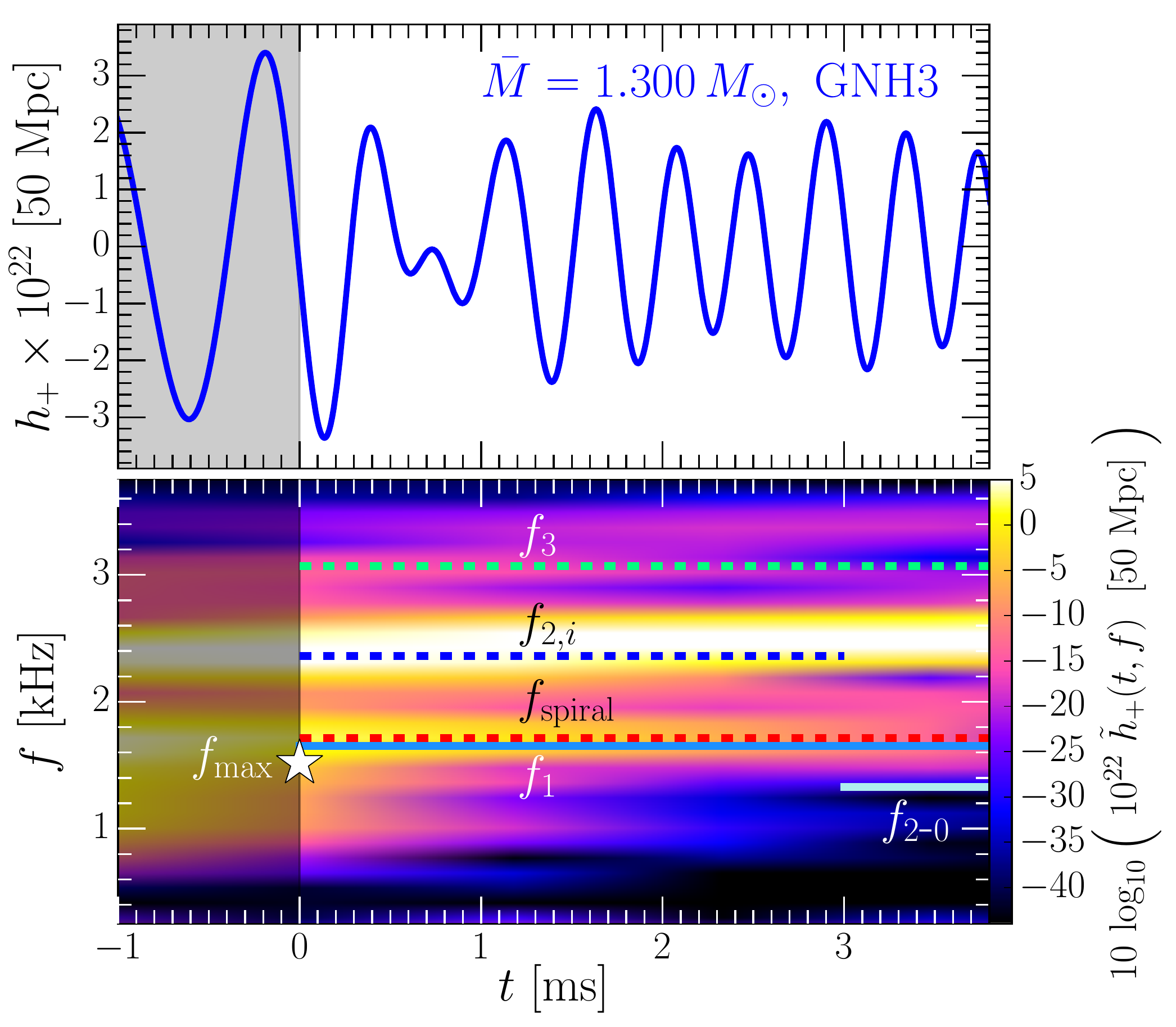}
\hskip 1.0cm
\includegraphics[width=7.5cm,height=7.5cm]
                {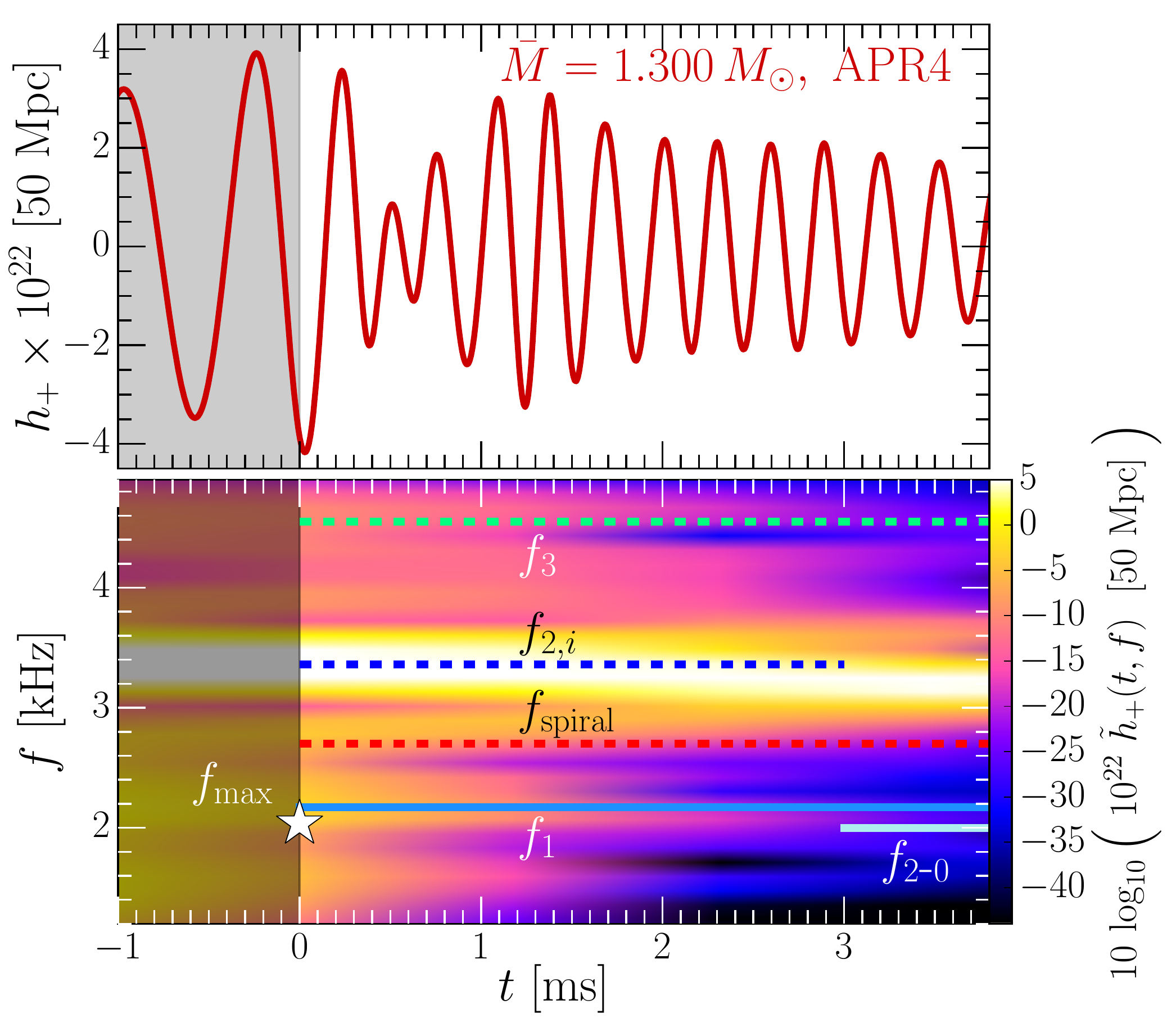}
\caption{Two examples of the GW emission around the merger, \ie $1\,{\rm
    ms}$ before the (gray-shaded area) and $4\,{\rm ms}$ after the
  merger. Both panels refer to a fiducial mass of $\bar{M}=1.300\,M_{\odot}$,
  with the left panel showing a representative stiff EOS (\ie GNH3),
  while the right panel a representative soft EOS (\ie APR4). The top
  part of each panel reports the gravitational strain $h_+$ for a source
  at $50\,{\rm Mpc}$, while the bottom part the corresponding
  spectrogram. Also marked with horizontal lines of different type and
  colour are the various frequencies discussed so far in the literature
  (see main text).}
\label{GWSpect_eg_sh}
\end{center}
\end{figure*}

\section{Results}
\label{sec:results}

\subsection{Waveform properties: transient signals}
\label{sec:wp_ts}

Figure \ref{Waveform_5x7} provides a summarising view of some of the
waveforms (\ie of $h_{+}$ for sources at a distance of $50\,{\rm Mpc}$)
computed in this paper and that are combined with those of
\cite{Takami2015} to offer a more comprehensive impression of the GW
signal across different masses and EOSs. The figure is composed of 35
panels referring to the 35 equal-mass binaries with nuclear-physics EOSs
that we have simulated and that have a postmerger signal of at least
$20\,{\rm ms}$; binaries with shorter post-merger (\eg
\mn{ALF2-q10-M1400}) and unequal-mass (\eg \mn{GNH3-q09-M1300}) are not
reported in the figure but their properties are listed in Table
\ref{tab:models}. Different rows refer to models with the same mass,
while different columns select the five cold EOSs considered and
colour-coded for convenience. It is then rather easy to see how small
differences across the various EOSs during the inspiral become marked
differences after the merger. In particular, it is straightforward to
observe how the GW signal increases considerably in frequency after the
merger and how low-mass binaries with stiff EOSs (\eg top-left panel for
the GNH3 EOS) show a qualitatively different behaviour from high-mass
binaries with soft EOSs (\eg bottom-right panel for the APR4 EOS). Also
quite apparent is that, independently of the mass considered, the
post-merger amplitude depends sensitively on the stiffness of the EOS,
with stiff EOSs (\eg GNH3) yielding systematically larger amplitudes than
soft EOSs (\eg APR4).

What is less evident from Fig. \ref{Waveform_5x7} are the features of the
transient GW signals emitted a few millisecond after the merger. To this
scope, we present in Fig. \ref{GWSpect_eg_sh} two representative examples
that concentrate on a time window around the merger, \ie one millisecond
in the inspiral (light-gray shaded area) and four milliseconds after the
merger. Both panels represent what could be the most realistic reference
value for the mass, \ie $\bar{M}=1.300\,M_{\odot}$, with the left panel
referring to a representative soft EOS (\ie GNH3), while the right panel
showing the same but for a representative stiff EOS (\ie APR4). The top
part of each panel reports the gravitational strain $h_+$ for a source at
$50\,{\rm Mpc}$, while the bottom part the corresponding spectrogram, \ie
the evolution of the PSD, where timeseries segments with length $\sim
8\,{\rm ms}$ and transformed with a Blackman window are overlapped by
$90\%$. Also marked with horizontal line of different type and colour are
the various frequencies that have been so far discussed in the literature
when describing the spectral properties of the GW signal.

Although we have already mentioned such spectral properties in the
Introduction, but we also briefly summarise them below:
\begin{itemize}

\item the frequencies $f_{\rm max}$ were first introduced in
  Ref. \cite{Read2013} and mark the instantaneous GW frequency at the
  merger, \ie at GW amplitude maximum. These frequencies were discussed
  in Ref. \cite{Read2013}, where they were first shown to correlate with
  the tidal deformability of the two stars; similar findings were later
  reported in Refs. \cite{Bernuzzi2014,Takami2015}. The values reported
  here are \emph{measured} from the data (see also Fig. \ref{Corr_Mfmax}
  below).

\item the frequencies $f_1, f_{2}, f_3$ were introduced in
  Refs. \cite{Takami2014,Takami2015} and represent the three main peaks
  of the PSDs measured in those references\footnote{Other authors, \eg
    \cite{Bauswein2011,Bauswein2015}, refer to the $f_2$ frequency as to
    $f_{\rm peak}$, but we find this convention confusing as there are
    several ``peaks'' in the PSD.}. The frequencies were found to roughly
  follow the relation $f_{2} \simeq (f_1 + f_3)/2$, and a simple
  mechanical toy model was proposed in \cite{Takami2015} to explain
  simply this relation.

\begin{figure*}
\begin{center}
\includegraphics[width=7.5cm,height=7.5cm]
                {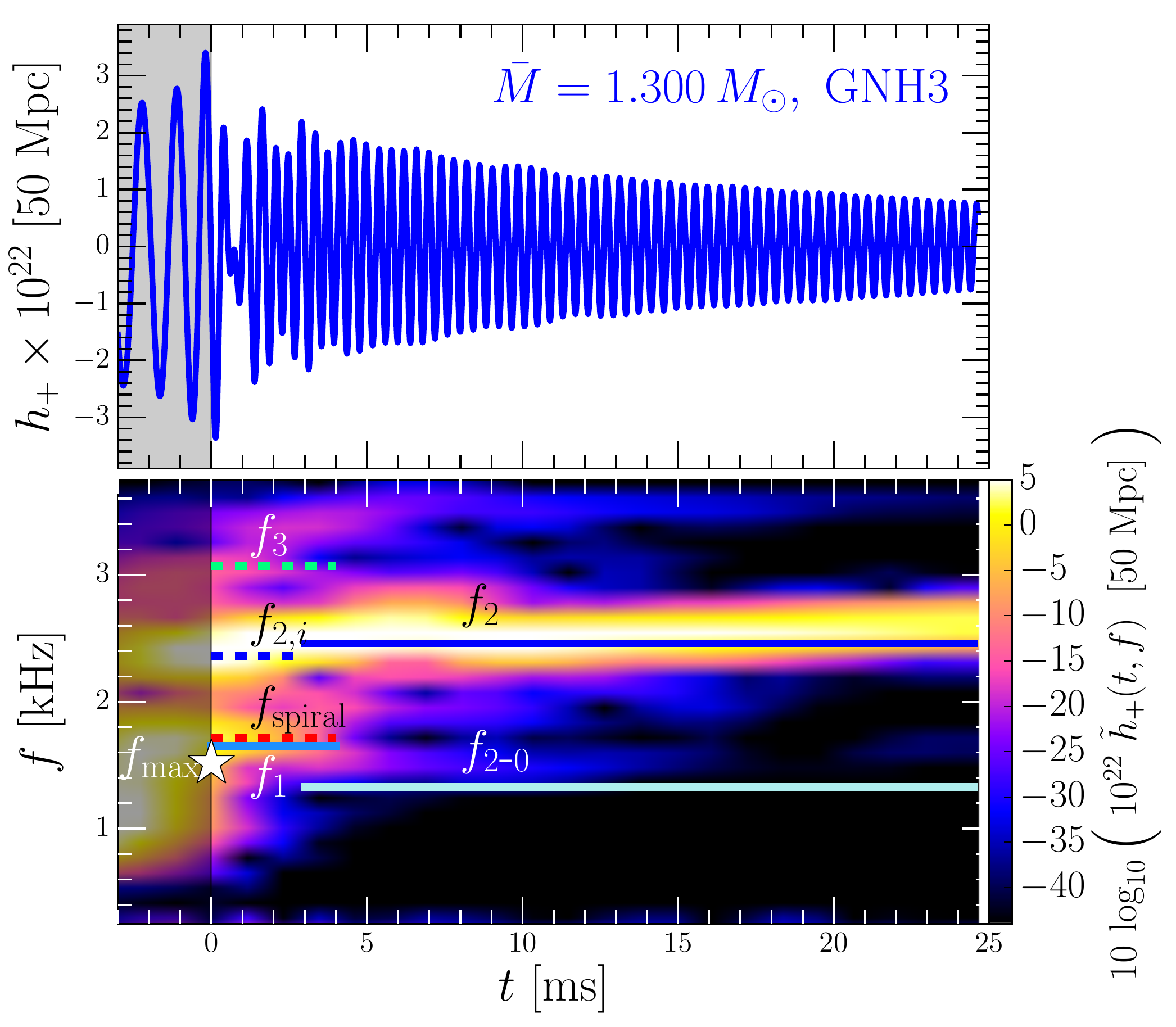}
\hskip 1.0cm
\includegraphics[width=7.5cm,height=7.5cm]
                {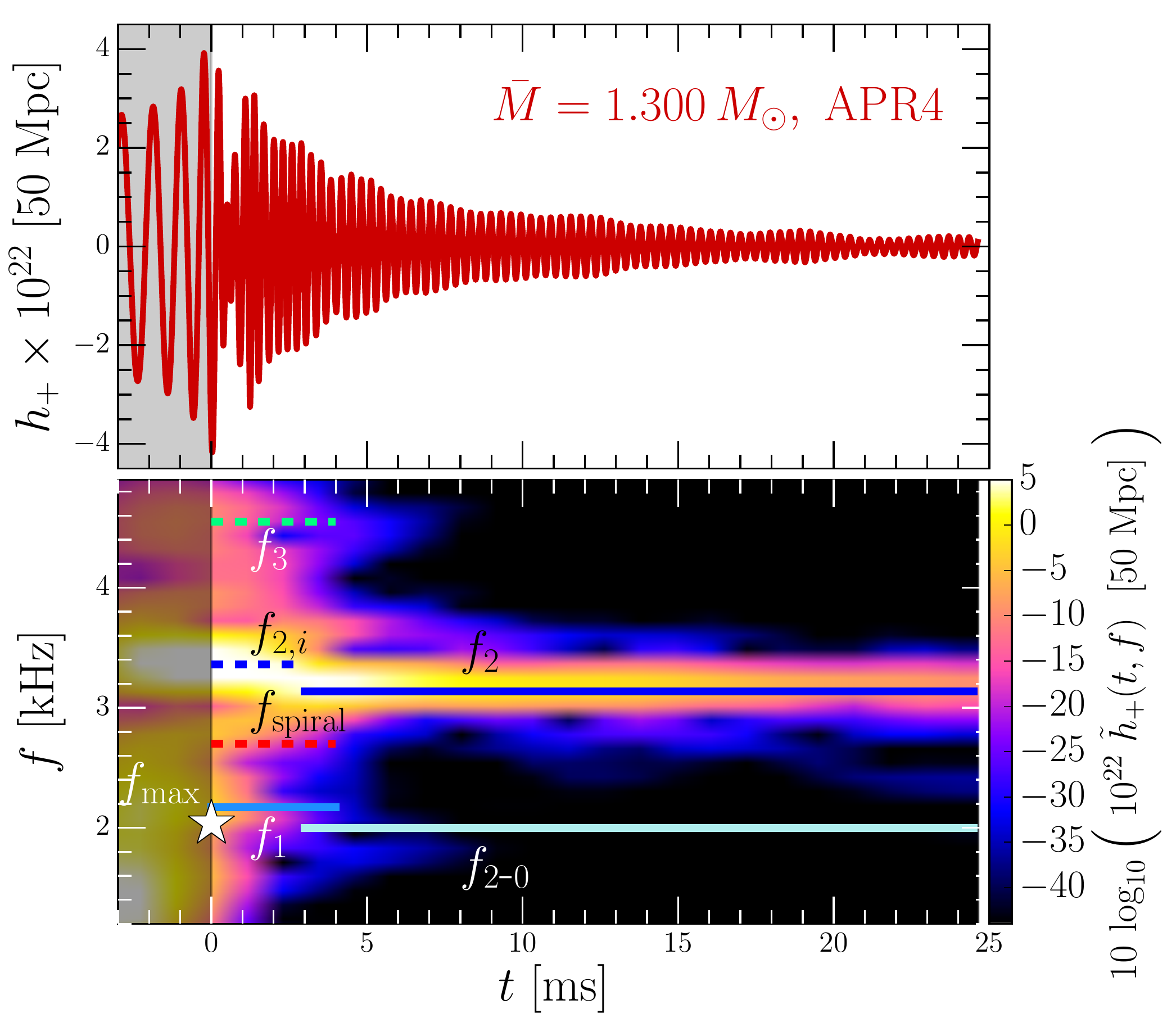}
\caption{The same as in Fig. \ref{GWSpect_eg_sh}, but shown on a much
  longer timescale, \ie $25\,{\rm ms}$ after the merger. Note that all of
  the peaks present in the short transient stage (\ie $t \lesssim 3\,{\rm
    ms}$) essentially disappear in the quasi-stationary evolution. The
  only exception is the $f_2$ peak, which slightly evolves from the
  $f_{2,i}$ frequency. In the case of a stiff EOS (\eg for the H4 EOS,
  but not shown here) a trace of the $f_{2\mbox{-}0}$ mode is still
  present, although at very low amplitudes.}
\label{GWSpect_eg_lg}
\end{center}
\end{figure*}

\item the $f_{2}$ frequencies correspond to the $\ell=2=m$ fundamental
  mode of the HMNS and hence are equal to twice the rotation frequency of
  the bar deformation of the HMNS. Their values change slightly in time
  (by $\sim 5\%$), and we indicate with $f_{2,i}$ the values in the
  \emph{transient} phase to distinguish them from the values $f_{2}$
  attained in the subsequent \emph{quasi-stationary} evolution of the GW
  signal.

\item the values reported here for $f_{2,i}, f_2$, and $f_1$ are
  \emph{measured} from the data. In particular, the $f_{2,i}$ frequencies
  are measured from the spectrograms, while the $f_2$ frequencies from
  the full PSDs; see Sect. \ref{sec:wp_aotfpsd}). For the $f_1$
  frequencies, instead, a first guess is \emph{predicted} from the
  analytic expression Eq. (25) of \cite{Takami2015} [\ie
    Eq. \eqref{eq:empirical_f1} here]; we then use these guesses and the
  spectrograms to refine the measured values of the $f_1$ frequencies,
  which we report in Figs. \ref{GWSpect_eg_sh}--\ref{PSD_5x7} and in
  Table \ref{tab:SNRs}\footnote{In pratice, we take the analytic guess
    from Eq. (25) of \cite{Takami2015} to draw a horizontal line in the
    spectrogram and then correct its vertical position of a few percent
    till it matches the largest value of the PSD for the longest amount
    of time.}. Such values of the $f_1$ frequencies are also used to
  obtain an improved estimate of the fitting coefficients
  \eqref{eq:empirical_f1_coeffs}. Finally, the third frequency is also
  \emph{predicted} as $f_3 = 2f_{2,i} - f_1$.

\item the frequencies $f_{2\mbox{-}0}$ were first introduced in
  Ref. \cite{Stergioulas2011b} and refer to a coupling between the
  $\ell=2=m$ fundamental mode and a quasi-radial axisymmetric mode, \ie
  with $m=0$, and that we indicate as $f_{m=0}$. The frequency of the
  latter mode can be \emph{measured}, for instance, from the oscillations
  in the lapse function or the rest-mass density at the center of the
  HMNS \cite{Bauswein2015,Bauswein2015b}, so that $f_{2\mbox{-}0} \equiv
  f_2 - f_{m=0}$ are effectively \emph{measured} quantities\footnote{Note
    that although the lapse function and the rest-mass density at the
    center of the HMNS are both a gauge quantities, they provide a robust
    and accurate representation of the eigenfrequnecies of oscillating
    compact objects, as shown in several studies
    \cite{Font99,Font02c,Baiotti04,Radice2011}.}.

\item the frequencies $f_{\rm spiral}$ were first introduced in
  Ref. \cite{Bauswein2015} and refer to the contribution to the GW signal
  coming from a ``rotating pattern of a deformation of spiral
  shape''. Because shapes in gauge-dependent quantities such as the
  rest-mass density are essentially impossible to measure in
  numerical-relativity calculations, we cannot measure these frequencies
  in our calculation. Hence, the values reported are those
  \emph{predicted} from the analytic prescription given in
  Ref. \cite{Bauswein2015} [\cf Eq. (2) of \cite{Bauswein2015}]. It was
  also claimed that the $f_{\rm spiral}$ peak can be roughly reproduced
  in a toy model, but no details were given in Ref. \cite{Bauswein2015}.
\end{itemize}
%

\begin{figure*}
\begin{center}
\includegraphics[width=8.75cm,height=13cm]
                {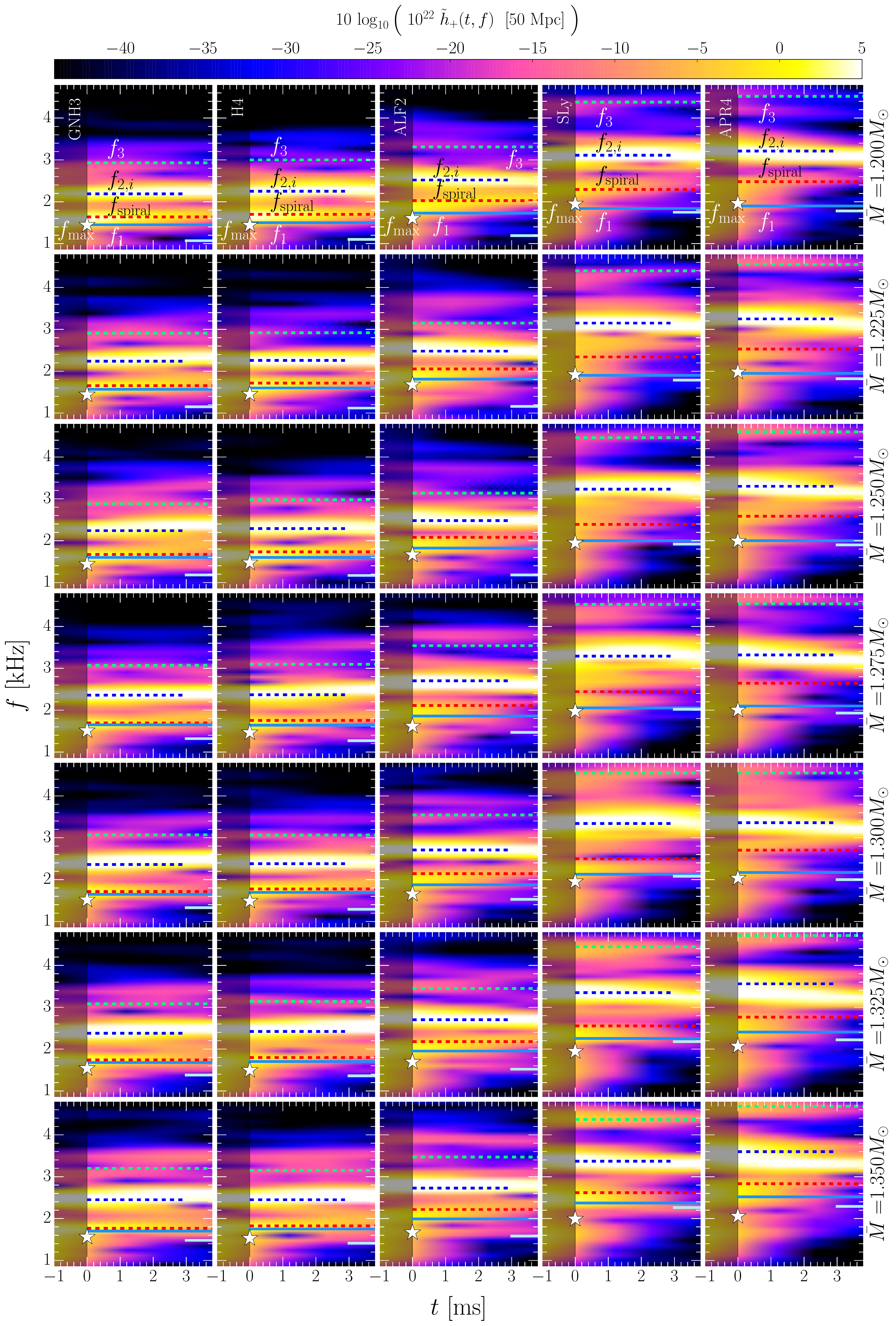}
\hskip 0.25cm
\includegraphics[width=8.75cm,height=13cm]
                {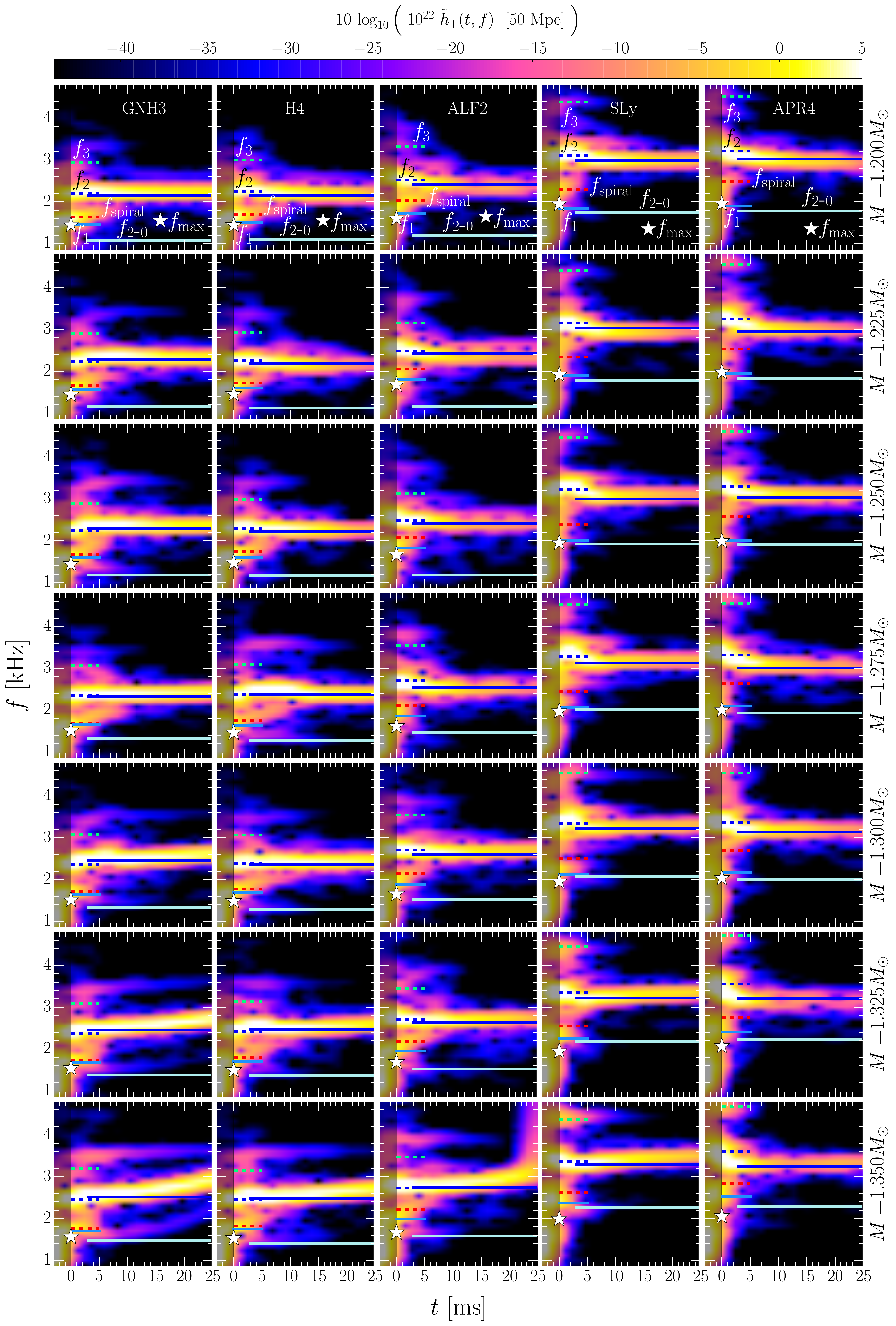}
\caption{The left and right panels show short and long spectrograms
  similar to those presented in Figs. \ref{GWSpect_eg_sh} and
  \ref{GWSpect_eg_lg}, but for the same set of binaries reported in
  Fig. \ref{Waveform_5x7}.}
\label{GWSpect_5x7}
\end{center}
\end{figure*}

The spectrograms in Fig. \ref{GWSpect_eg_sh} contain a wealth of
information about the transient post-merger phase. To clarify a series of
imprecise and sometimes confusing statements recently appeared in the
literature, we collect below the main information on the spectral
properties of the \emph{transient}. The points summarised below do not
refer only to the models in Figs. \ref{GWSpect_eg_sh} and
\ref{GWSpect_eg_lg}, but to all of the models simulated [\cf
  Fig. \ref{GWSpect_5x7}].

\begin{figure*}
\begin{center}
\includegraphics[width=2.0\columnwidth]{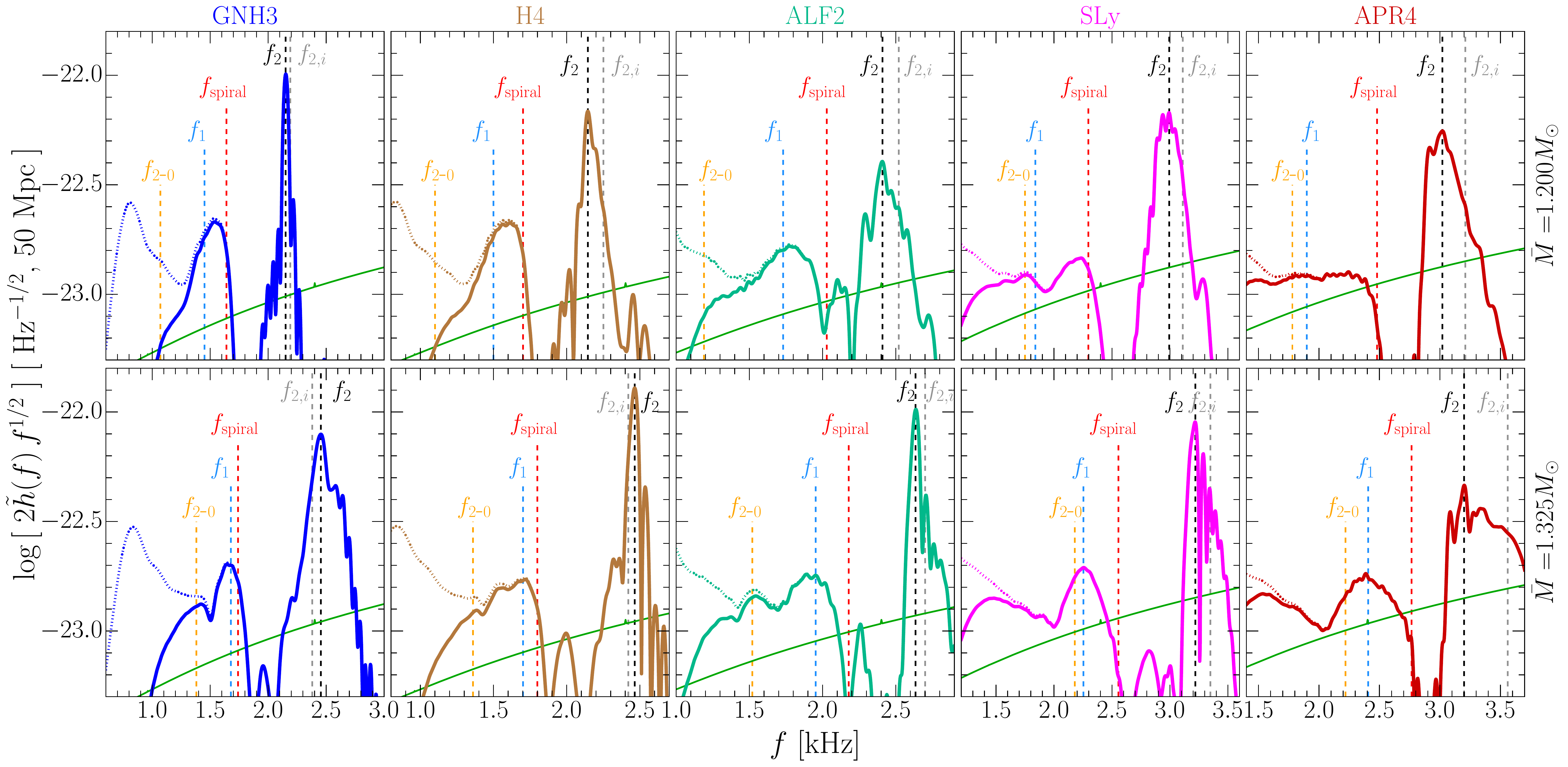}
\caption{Total PSDs of the GW signals for all the EOSs considered and
  relative to a series of low-mass (\ie $\bar{M}=1.200\,M_{\odot}$) and
  medium-mass binaries (\ie $\bar{M}=1.325\,M_{\odot}$); these are the
  two masses also considered in Ref. \cite{Bauswein2015}. The solid lines
  refer to the post-merger signal only, while the dotted lines report
  also the power during the short inspiral. Reported with vertical dashed
  lines of different colours are instead the values of the frequencies
  $f_{1}$, $f_{2,i}$, $f_{2}$, $f_{2\mbox{-}0}$, and $f_{\rm
    spiral}$. These are either measured from the PSDs or estimated
  numerically (see discussion in Sect. \ref{sec:wp_ts}). The PSDs are
  relative to GWs from binaries at a distance of $50\,{\rm Mpc}$ and we
  report also the sensitivity curves of Advanced LIGO as a reference
  (green line). Note that the $f_1$ and $f_{\rm spiral}$ frequencies
  (blue and red vertical dashed lines) are very similar for stiff EOSs
  (\eg GNH3, H4), but significantly different for soft EOSs (\eg SLy,
  APR4), making a proper distinction fallible. Note also that the
  $f_{2\mbox{-}0}$ frequencies (orange vertical dashed lines) do not
  always correspond to clearly identifiable peaks in the total PSDs.}
\label{PSD_2x5}
\end{center}
\end{figure*}

\begin{itemize}

\item the frequencies $f_{2,i}$ (blue dashed lines) are
  \emph{short-lived} and evolve into the $f_2$ frequencies as the GW
  signal reaches its quasi-stationary phase (\ie the one after $t\simeq 3
  \,{\rm ms}$). This change is of the order of $\sim 5\%$, so that $f_{2}
  \sim (1 \pm 0.05)\,f_{2,i}$.

\item the frequencies $f_1, f_3$ (light blue solid lines and green dashed
  lines) are instead \emph{short-lived} and their amplitude becomes
  vanishingly small after the transient. This was remarked in
  Ref. \cite{Takami2015}, where a simple toy model was developed to
  explain these frequencies as a result of the modulation of the
  (rotating) oscillation of the two stellar cores [\cf Appendix A of
    \cite{Takami2015}]. Note that although $f_1, f_3$ are only
  \emph{predicted} analytically, they do coincide with the maximum values
  of the spectrogram. This provides an important confirmation on the
  correctness of the interpretation in Ref. \cite{Takami2015}.

\item the frequencies $f_{2\mbox{-}0}$ (cyan solid lines) do not have
  significant power during the transient phase and it is only later that
  they may produce a contribution and only for rather stiff EOSs (\cf
  Fig. \ref{GWSpect_5x7}, which we discuss below). Note also that the
  frequencies $f_{2\mbox{-}0}$ are \emph{measured} from $f_{2}$ and
  $f_{m=0}$, respectively, leaving no room for interpretation.

\item the frequency $f_{\rm spiral}$ (red dashed lines) is essentially
  the \emph{same} as the $f_1$ frequency for the stiff EOS GNH3, while it
  is significantly \emph{different} for the soft EOS APR4. The fact that
  $f_{\rm spiral} \sim f_1$ in some cases but not in all cases, holds
  true also when considering other EOSs and was found also in
  Ref. \cite{Bauswein2015}. We will touch on this point further below (see
  dashed blue and red lines in Figs. \ref{PSD_2x5} and \ref{PSD_5x7} and
  relative discussion).

\item because the frequency $f_{\rm spiral}$ is \emph{predicted} from an
  analytic expression [\cf Eq. (2) of \cite{Bauswein2015}], its
  coincidence with the $f_1$ frequency for stiff EOSs suggests that the
  two frequencies are just the same for stiff EOSs. On the other hand,
  for soft EOSs the frequencies $f_{\rm spiral}$ do select a genuinely
  different mode, which are however \emph{short-lived}; as we will show
  in Fig. \ref{PSD_2x5}, the total power stored in these $f_{\rm spiral}$
  modes is always very small in our data (see also \cite{Bauswein2015}).

\item the only partial correspondence between the $f_1$ and $f_{\rm
  spiral}$ frequencies can explain why the $f_1$ frequencies are found
  to behave universally \cite{Takami2014, Takami2015}, while the $f_{\rm
    spiral}$ are not \cite{Bauswein2015}.

\end{itemize}

To recap, the analysis of the spectrograms in the transient phase reveals
that three modes are clearly visible: $f_{2,i}, f_1, f_3$. The last two
disappear later on, while the first one survives as $f_2$ and with
changes of a few percent. The $f_{\rm spiral}$ frequencies essentially
coincide with the $f_{1}$ frequencies for stiff EOSs, while marking a
different mode for soft EOSs; these latter frequencies are short lived
and provide a minimal contribution to the total PSD.

\subsection{Waveform properties: quasi-stationary signals}
\label{sec:wp_qss}

We next discuss the spectral properties of the signal when considered
over a much longer timescale, which we take to be at least $20\,{\rm ms}$
after the merger if the HMNS does not collapse before. This is shown in
Fig. \ref{GWSpect_eg_lg}, again for two representative EOSs and for the
fiducial mass of $\bar{M}=1.300\,M_{\odot}$. As anticipated in the previous
Section, the only frequency surviving on these timescales is the $f_2$
peak (blue solid lines), which evolves slightly from the $f_{2,i}$ peak
as the HMNS attains a quasi-stationary equilibrium. Other peaks, such
those associated to $f_1, f_3, f_{\rm spiral}$ essentially vanish after
the transient, while the $f_{2\mbox{-}0}$ peak retain only small powers.

Figure \ref{GWSpect_5x7} is rather ``dense'', but provides a
comprehensive summary in terms of spectrograms of the results discussed
in the last two sections. In particular, the left panel reports the
spectrograms for the 35 binaries presented in Fig. \ref{Waveform_5x7},
but concentrating on the transient phase, \ie for $t\in [-1,4]\,{\rm
  ms}$, while the right panel shows the spectrograms for the complete GW
signal. In essence, the two panels show that:

\begin{itemize}

\item for all the EOSs considered here, three frequencies appear in the
  transient phase: $f_1, f_{2,i}, f_3$. 

\item the $f_{\rm spiral}$ frequencies essentially coincide with the
  $f_{1}$ frequencies for stiff EOSs (\ie GNH3, H4, and ALF2), but differ
  for soft EOSs (\ie SLy and APR4).

\item for soft EOSs, the $f_{\rm spiral}$ frequencies are systematically
  at larger frequencies than the $f_{1}$ frequencies (see, \eg the model
  with $\bar{M}=1.225\,M_{\odot}$ for the APR4 EOS), but yield a very
  contribution to the overall PSD.

\item after the transient, only the $f_2$ frequencies survive as an
  adjustment of the $f_{2,i}$ frequencies produced during the transient.

\item the $f_{2\mbox{-}0}$ frequency can be easily measured from the
  oscillations of the lapse function but the associated power in the
  spectrograms is always extremely small and appreciable only after the
  transient and for a limited period of time (see, \eg model with
  $\bar{M}=1.300-1.350\,M_{\odot}$ for stiff EOSs GNH3, H4 and ALF2).

\end{itemize}

\subsection{Waveform properties: analysis of the full PSDs}
\label{sec:wp_aotfpsd}

The discussion has so far been focused on the analysis of the spectral
properties of the GW signal as deduced when looking at the
spectrograms. We next investigate how the spectral properties appear when
analysing the full PSDs of the GW signal.

We start this discussion by considering two representative examples in
Fig. \ref{PSD_2x5}, which reports the PSDs of a series of low-mass (\ie
$\bar{M}=1.200\,M_{\odot}$) and medium-mass binaries (\ie
$\bar{M}=1.325\,M_{\odot}$); solid lines refer to the post-merger signal
only, while the dotted lines report also the power during the short
inspiral. The frequencies relative to the peaks $f_{1}$, $f_{2,i}$,
$f_{2}$, and $f_{2\mbox{-}0}$ are shown with vertical dashed lines of
different colours. Additionally, the $f_{\rm spiral}$ frequency of
Ref. \cite{Bauswein2015}, which was reported for
$\bar{M}=1.200,~1.350$ and $1.500\,M_{\odot}$ [\cf Eq.\,(2)
  of \cite{Bauswein2015}], are also shown as a reference. In essence, the
PSDs in Fig. \ref{PSD_2x5} reveal that:

\begin{figure*}
\begin{center}
\includegraphics[width=16.0cm,height=18cm]{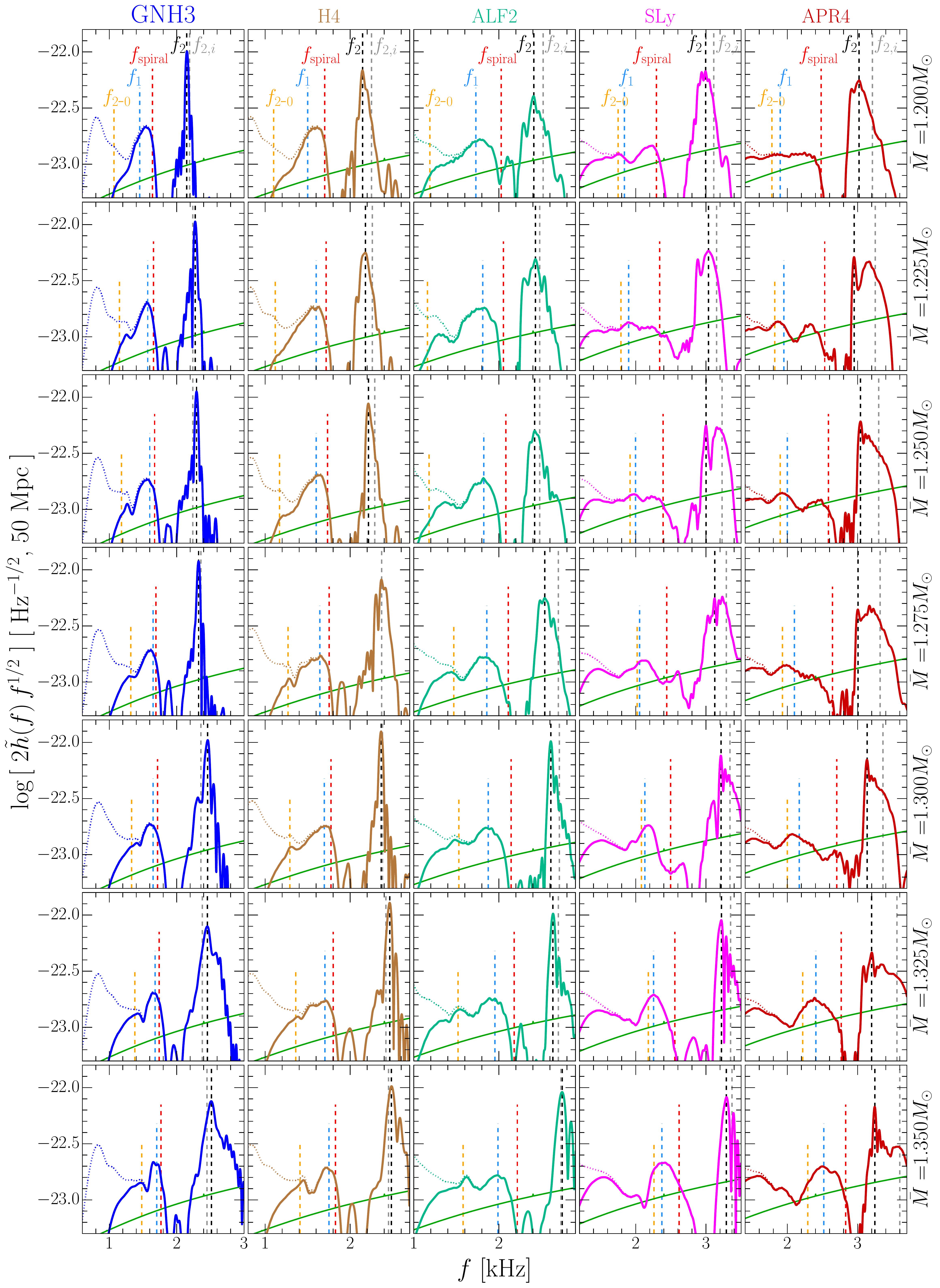}
\caption{The same as in Fig.~\ref{PSD_2x5}, but for the same set of
  binaries reported in Fig. \ref{Waveform_5x7}. Note that the $f_1$ and
  $f_{\rm spiral}$ frequencies (blue and red vertical dashed lines) are
  very similar for stiff EOSs (\eg GNH3, H4), but significantly different
  for soft EOSs (\eg SLy, APR4).}
\label{PSD_5x7} 
\end{center}
\end{figure*}

\begin{figure*}
\begin{center}
\includegraphics[width=0.85\columnwidth]{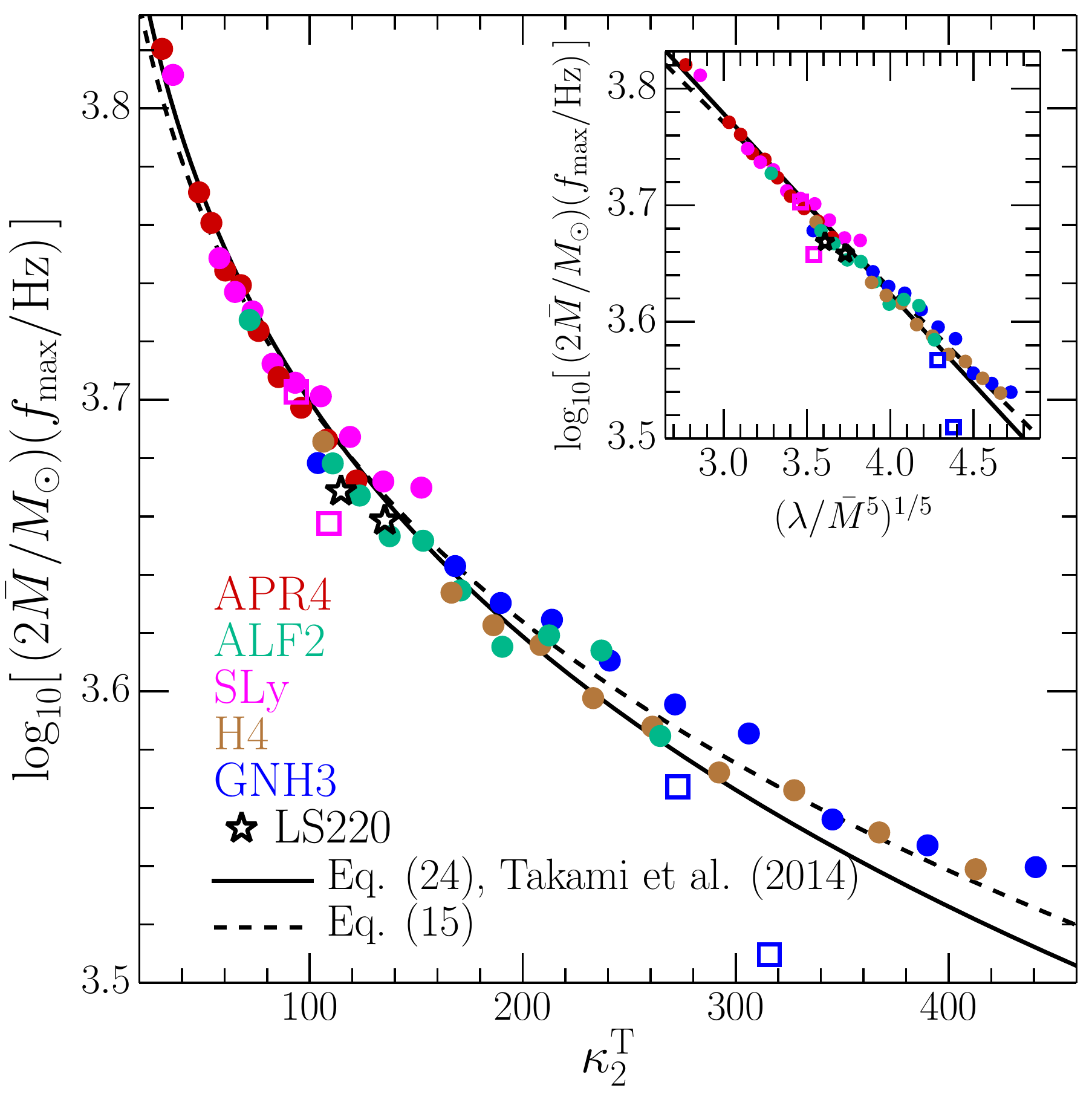}
\hskip 2.0cm
\includegraphics[width=0.85\columnwidth]{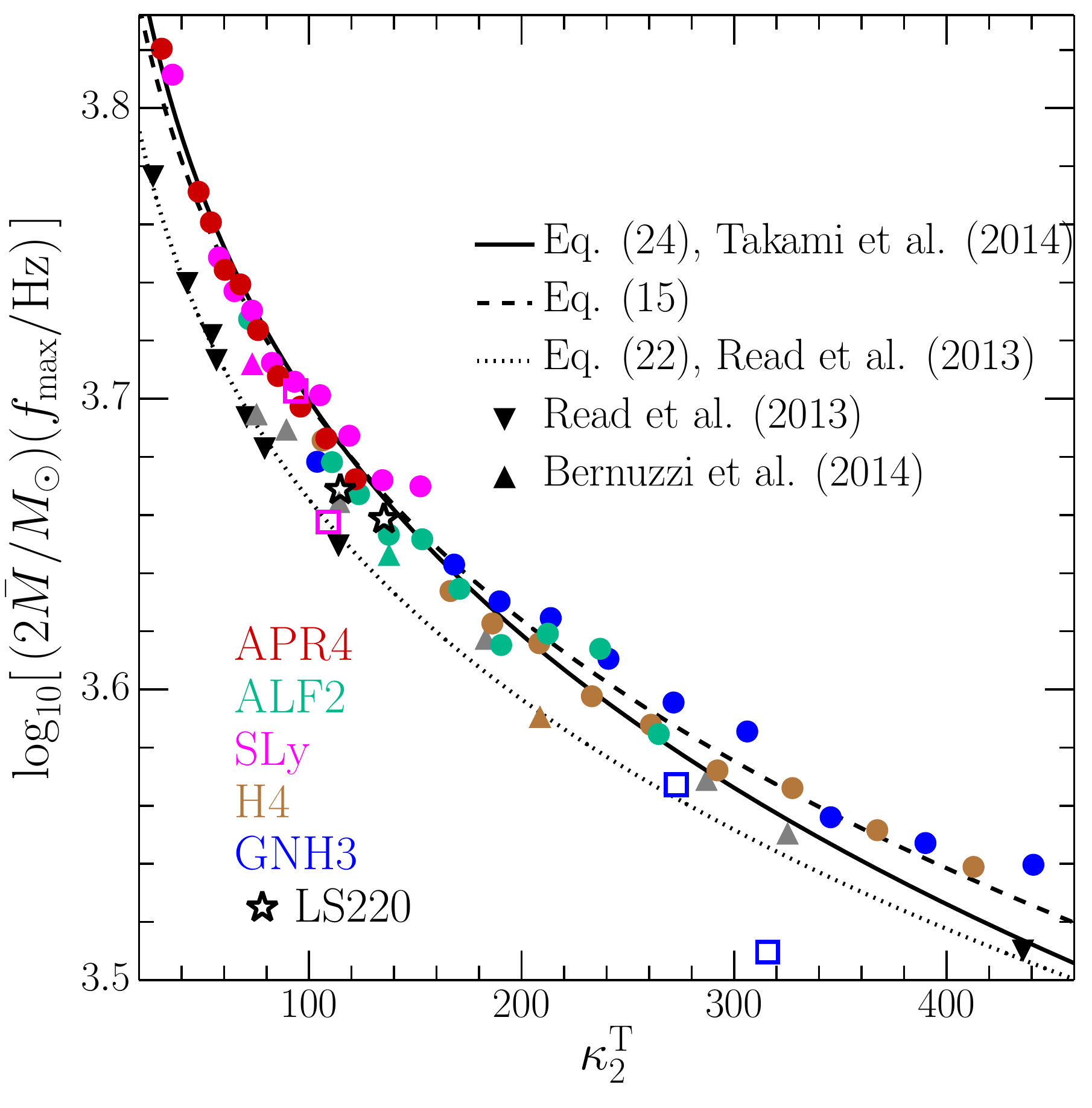}
\caption{\textit{Left panel:} Mass-weighted frequencies at amplitude
  maximum $f_\mathrm{max}$ shown as a function of the dimensionless tidal
  deformability $\kappa^{^T}_2$ (the inset shows the same data in terms
  of $(\lambda/\bar{M}^5)^{1/5}$ to highlight the essentially linear
  behaviour). Filled circles of different colours refer to equal-mass
  binaries with different cold EOSs, the empty squares to the
  unequal-mass binaries, and the stars to the hot-EOS binariess. The
  black solid line shows the fit as given by Eq.~(24) of
  \cite{Takami2015}, while the black dashed line the fit with the updated
  coefficients \eqref{eq:coeff_fmax_new}. \textit{Right panel:} The same
  as in the left panel but showing also the data from
  Refs. \cite{Read2013,Bernuzzi2014}, where a slightly different
  definition of the merger time is used and leads to the larger
  variance.}
\label{Corr_Mfmax} 
\end{center}
\end{figure*}

\begin{itemize}

\item for all the EOSs considered, the peak corresponding to the $f_2$
  frequency is rather easy to recognise and is reasonably well reproduced
  by an analytic expression that we will discuss in
  Sect. \ref{sec:wp_cwsp}.

\item the $f_{2,i}$ frequencies as measured from the spectrograms do not
  correspond to any visible peak in the total PSDs; this is to be
  expected given that these frequencies are only short lived and their
  contribution to the total PSD is much smaller than that of the $f_2$
  frequencies.

\item smaller but still clearly visible are the contributions of the
  $f_1$ and $f_3$ frequencies (the latter are not reported in
  Fig. \ref{PSD_2x5} for clarity). This behaviour too is not surprising
  and is due to the short duration of these modes. Note that the $f_1$
  frequencies in Fig. \ref{PSD_2x5}, which we recall are \emph{predicted}
  analytically, also mark the presence of a local maximum in the PSD.

\item for stiff EOSs, \eg GNH4 and H4, the peaks corresponding to
  the $f_1$ and $f_{\rm spiral}$ frequencies are very similar, but they
  become distinct for soft EOSs, \eg ALF2, SLy and APR4.

\item when not being comparable to $f_1$, the $f_{\rm spiral}$
  frequencies do not seem to mark any local maximum in the PSDs, see, for
  example the BNS with $\bar{M}=1.200\,M_{\odot}$ and the SLy EOS, or the
  BNS with $\bar{M}=1.350\,M_{\odot}$ and the EOSs SLy and APR4.

\item the behaviour of the $f_{2\mbox{-}0}$ frequencies is far less
  clear. In those cases where it is comparable with the $f_1$ frequencies
  (\eg for the SLy EOS), these frequencies can be associated to the same
  power excess attributed to the $f_1$ frequencies. In other cases,
  however, they are either associated to peaks with very limited
  power\footnote{This is the case, for instance, for the binaries with
    $\bar{M}=1.350\,M_{\odot}$ with EOS GNH3, H4 and ALF2.} or are
  associated to peaks\footnote{This is the case, for instance, for the
    binaries with $\bar{M}=1.200\,M_{\odot}$ with EOS GNH3, H4 and ALF2,
    or for the binaries with $\bar{M}=1.350\,M_{\odot}$ with EOS
    APR4.}. This is not surprising since the $f_{2\mbox{-}0}$ peaks
  result from a mode coupling and are therefore expected to be less
  energetic.

\end{itemize}

The properties of the PSDs listed above and illustrated in
Fig. \ref{PSD_2x5} are not limited to the cases of the binary masses
reported in that figure. This conclusion can be reached after inspecting
Fig. \ref{PSD_5x7}, which is the same as Fig. \ref{PSD_2x5}, but reports
also all the other masses considered. For compactness we do not report
here the PSDs relative to the unequal-mass binaries considered, and whose
PSDs show a very similar behaviour to the ones discussed so far.

\subsection{Waveform properties: correlations with the stellar properties}
\label{sec:wp_cwsp}

In what follows we make use of the results discussed so far to correlate
the spectral properties of the GW signal with the properties of the
progenitor stellar models. As discussed extensively in
Refs. \cite{Bauswein2011, Bauswein2012, Read2013, Bernuzzi2014,
  Takami2014, Takami2015, Bernuzzi2015a}, some of these correlations
appear to be ``universal'', \ie only slightly dependent on the EOS, and
can therefore be used to constrain the physical properties of the
progenitor stars and hence the EOS.

\subsubsection{Inspiral and merger}

We start by considering the \emph{inspiral} part of the signal. Figure
\ref{Corr_Mfmax} reports the mass-weighted frequencies at amplitude
maximum $f_\mathrm{max}$ as a function of the tidal polarizability
parameter $\kappa_2^{^T}$ for a generic unequal-mass binary. We recall
that the latter is defined as (see, \eg \cite{Bernuzzi2014})
\begin{equation}
\label{kappa_Bernuzzi}
\kappa_2^{^T} \equiv
2\left[
         q \left(\frac{X_{_A}}{C_{_A}}\right)^5k^{^A}_2 + 
\frac{1}{q}\left(\frac{X_{_B}}{C_{_B}}\right)^5k^{^B}_2\right]\,,
\end{equation}
where $A$ and $B$ refer to the primary and secondary stars in the binary
\begin{align}
&q \equiv \frac{M_{_B}}{M_{_A}} \leq 1\,, &
&X_{_{A,B}} \equiv \frac{M_{_{A,B}}}{M_{_A}+M_{_B}}\,, &
\end{align}
$k_2^{^{A,B}}$ are the $\ell=2$ dimensionless tidal Love numbers, and
$\mathcal{C}_{_{A,B}} \equiv M_{_{A,B}}/R_{_{A,B}}$ are the
compactnesses. In the case of equal-mass binaries, $k_2^{A}=k_2^{B}=\bar{k}_2$,
and expression \eqref{kappa_Bernuzzi} reduces to
\begin{equation}
\kappa_2^{^T} \equiv \frac{1}{8}\bar{k}_2
\left(\frac{\bar{R}}{\bar{M}}\right)^5 = \frac{3}{16}\Lambda =
\frac{3}{16} \frac{\lambda}{\bar{M}^5} \,,
\end{equation}
where the quantity
\begin{equation}
\lambda \equiv \frac{2}{3} \bar{k}_2 {\bar{R}}^5\,.
\end{equation}
is another commonly employed way of expressing the tidal Love number for
equal-mass binaries \cite{Read2013}, while $\Lambda \equiv
\lambda/\bar{M}^5$ is its dimensionless counterpart and was employed in
\cite{Takami2015}.

The black solid line in the left panel of Fig. \ref{Corr_Mfmax} shows the
fit as given by Eq.~(24) of \cite{Takami2015}, where five realistic
EOSs~(\ie APR4, ALF2, SLy, H4, GNH3) and an ideal-fluid EOS with
$\Gamma=2$ were employed across a range of five different masses for each
EOS. Such a fit was expressed as [\cf Eq.~(24) of \cite{Takami2015}]
\begin{equation}
\label{eq:oldfit}
\log_{10} \left(\frac{f_\mathrm{max}}{\mathrm{kHz}}\right) \approx
 a_0  + a_1\, \left(\kappa^{^T}_2\right)^{\!\!1/5} \!- \log_{10}
 \left(\frac{2\bar{M}}{\Msun}\right)
\,,
\end{equation}
where\footnote{Note that in \cite{Takami2015} the fit was actually done
  using the quantity ${\lambda}/{\bar{M}^5}$, so that the coefficient
  $a_1$ reported there is different by a factor $(16/3)^{1/5}$.}
\begin{align}
\label{eq:coeff_fmax_old}
&a_0 = 4.242\,, &a_1= -0.216\,. 
\end{align}
Making use of the larger sample of binaries and excluding from the fit
the ideal-fluid EOS, we can further refine the fit and obtain new and
slightly modified coefficients
\begin{align}
\label{eq:coeff_fmax_new}
&a_0 = 4.186\,, &a_1= -0.195\,. 
\end{align}
The new fit is indicated with a black dashed in the left panel of
Fig. \ref{Corr_Mfmax}, which also reports in an inset the same data but
when represented via the dimensionless tidal deformability
$(\lambda/M^5)^{1/5}$ to highlight the essentially linear dependence in
terms of this variable.

In the right panel of Fig. \ref{Corr_Mfmax}, we instead show the same
correlation as in the left panel (and the corresponding fitting
expressions) but when using the data taken from Refs. \cite{Read2013,
  Bernuzzi2014}. Note that in these cases, the frequencies reported are
measured at the peak of amplitude of one of the polarization modes of the
strain, \ie at the maximum amplitude of $h^2_+(t)$ rather than of $h(t)
\equiv \left(h^2_+ + h^2_\times\right)^{1/2}$; we believe this slight
differences is responsible for the larger variance in the correlation 
(see discussion in Sect. V C. in \cite{Takami2015}).

\begin{figure*}
\begin{center}
\includegraphics[width=0.85\columnwidth]
                {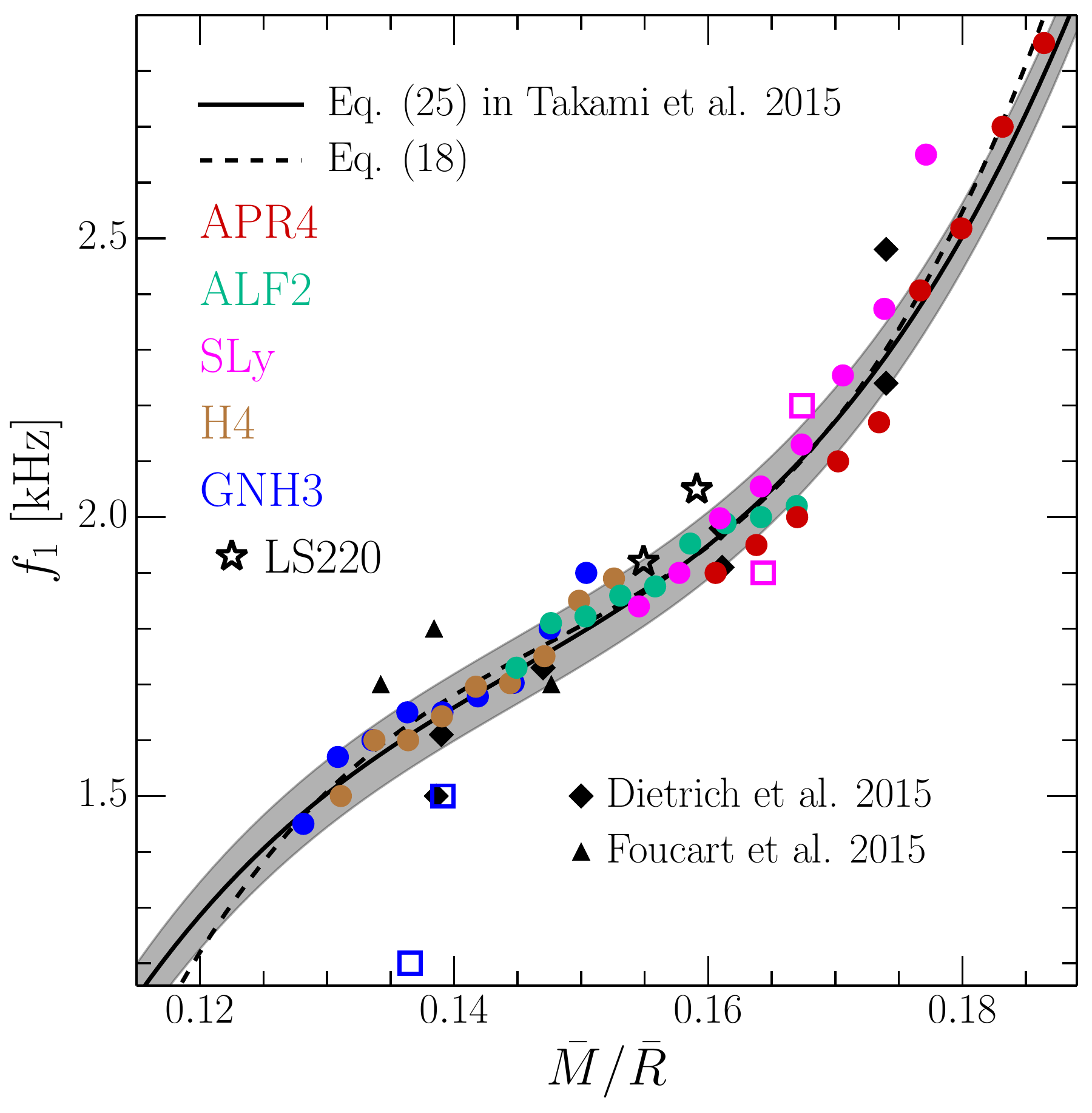}
\hskip 2.0cm
\includegraphics[width=0.85\columnwidth]
                {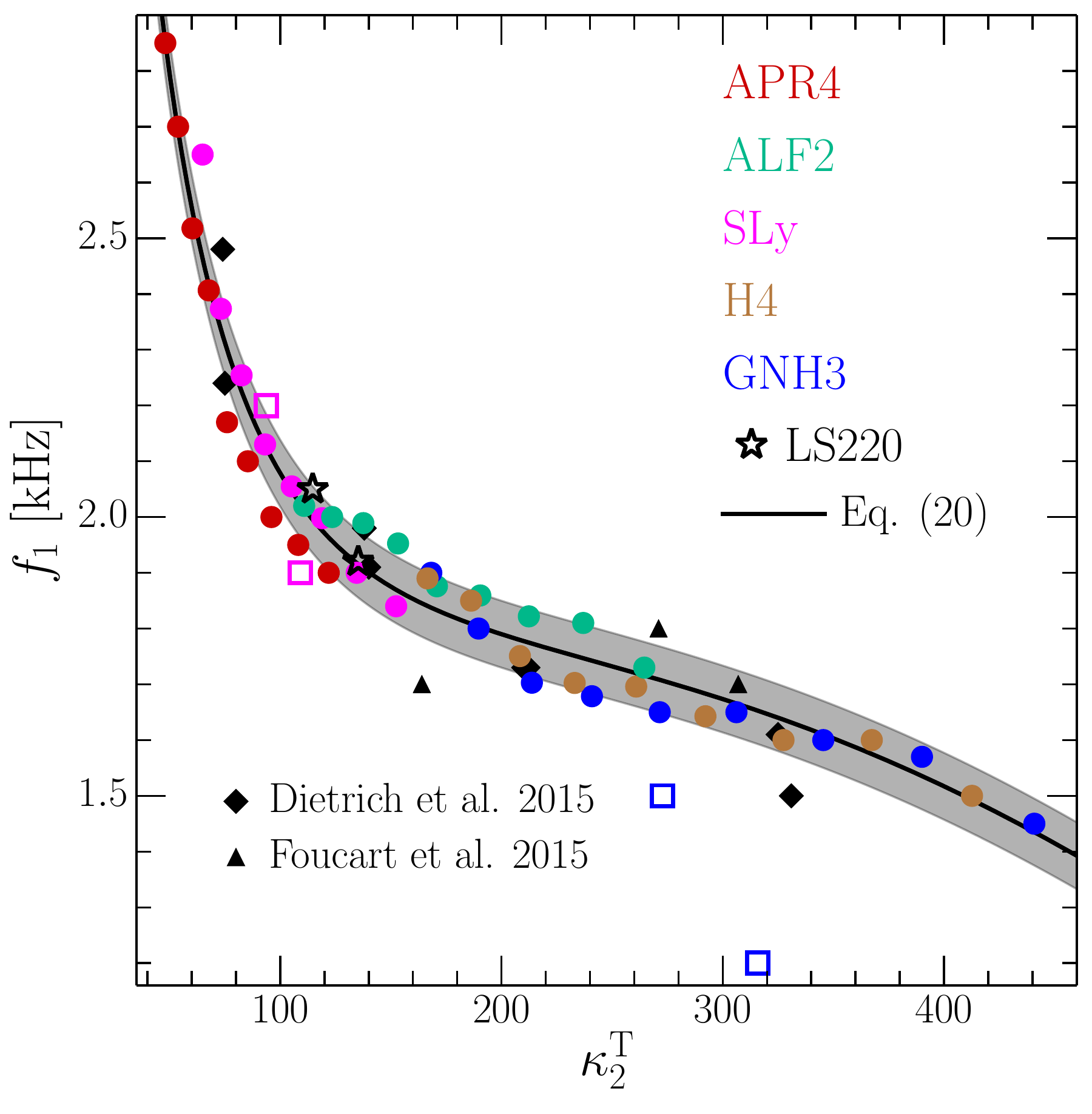}
\caption{\textit{Left panel:} Values of the low-frequency peaks $f_1$ for
  equal-mass binaries (we use the same symbol convention as in
  Fig. \ref{Corr_Mfmax}) shown as a function of the average stellar
  compactness. The solid black line shows the quasi-universal relation
  first reported in \cite{Takami2015}. Note that the new data, including
  the low- and high-mass models (\eg $\bar{M}=1.200, 1.500\,\Msun$), respect
  the ``universal'' behaviour. Also reported is the data presented in
  Ref. \cite{Dietrich2015} (black diamonds) and \cite{Foucart2015} (black
  triangles). \textit{Right panel:} The same data as in the left panel,
  but shown as a function of the dimensionless tidal deformability
  $\kappa^{^T}_2$ and highlighting a quasi-universal relation also in
  terms of this quantity.}
\label{Corr_f1} 
\end{center}
\end{figure*}

Overall, the data reported in Fig. \ref{Corr_Mfmax} confirms what was
first pointed out in Ref. \cite{Read2013}, namely, that a rather tight
``universal'' correlation exists between the frequency at peak amplitude
and the tidal deformability; for equal-mass binaries, the largest
difference between the values measured for $f_{\rm max}$ and those
predicted by the fit are $\simeq 3.6\%$, but the average deviation is
much smaller and $\simeq 1.3\%$ only.

The correlation becomes weaker when considering unequal-mass binaries and
this is clearly shown by the data marked with empty squares at
$\kappa^{^T}_2 \simeq 105$ and at $\kappa^{^T}_2 \simeq 315$. Both points
refer respectively to the APR4 and GNH3 binaries with the smallest mass
ratio of $q\simeq 0.8$ and this can be interpreted as the ``breaking'' of
the universality for small (and possibly unrealistic) mass ratios. Given
that the dynamics and GW signal in these cases is rather different from
the corresponding equal-mass binaries (the merger necessarily happens at
lower frequencies as the tidal interaction is amplified and the
lower-mass star disrupted), this is perfectly plausible; however
additional simulations will be needed to confirm this conjecture.

\subsubsection{Post-merger}

Next, we consider the correlations in the \emph{post-merger} part of the
signal and concentrate initially on the low-frequency $f_1$ peaks. The
left panel of Fig. \ref{Corr_f1} reports the values of such frequencies
as a function of the stellar compactness. The solid black line represents
the ``universal'' relation first reported in Ref. \cite{Takami2015} and
expressed as a cubic polynomial [\cf Eq.~(25) of Ref. \cite{Takami2015}]
\begin{equation}
f_1 \approx b_0 + b_1~{\mathcal C} +
b_2~{\mathcal C}^2 + b_3~{\mathcal C}^3\ {\mathrm{kHz}}\,,
\label{eq:empirical_f1}
\end{equation}
where the fitting coefficients given by \cite{Takami2015} have been
further refined after using the larger sample of data considered here and
are given by
\begin{align}
&b_0 = -35.17    \,,& 
&b_1 = 727.99    \,, &\nonumber\\
&b_2 = -4858.54  \,,& 
&b_3 = 10989.88  \,.
\label{eq:empirical_f1_coeffs}
\end{align}
Also indicated with a gray shaded band is the overall $\lesssim
0.06\,{\rm kHz}$ uncertainty in the fit related either to the truncation
error or to the determination of the frequencies \cite{Takami2015}.

Note that the data relative to the new binaries is in full agreement with
our previous results in \cite{Takami2015} and that even the very low-mass
binaries, \ie with $\bar{M}=1.200\,\Msun$, support the universal
relation. Hence, we believe that the mismatch found in
\cite{Bauswein2015} is due to the incorrect association of the $f_1$
frequency with the $f_{\rm spiral}$ frequencies. 

Also reported in the figure are the putative $f_1$ frequencies as taken
from Refs. \cite{Dietrich2015, Foucart2015}, respectively for binaries
with $\bar{M}=1.350\,\Msun$ and $q=0.862$ and $q=1$, evolved with EOSs
MS1, H4, ALF2, and SLy \cite{Dietrich2015}, as well as for binaries with
$\bar{M}=1.200\,\Msun$ and $q=1$ evolved with EOSs LS220, DD2, and SFHo
\cite{Foucart2015}. In particular, in the case of the SFHo EOS, we
believe that the value reported in Table II of \cite{Foucart2015} (\ie
$2.1\,{\rm kHz}$) actually refers to the $f_{\rm spiral}$ frequency and
that the correct value for the $f_1$ frequency is the low-frequency one
which is clearly marked in the left panel of Fig. 6 in \cite{Foucart2015}
(\ie $1.7\,{\rm kHz}$). We have already remarked that in the case of soft
EOSs and low-mass binaries it is very easy to confuse the $f_1$ frequency
with the $f_{\rm spiral}$ one; we believe this is one of those
instances. Indeed, when this correction is made, all of the data of
Refs. \cite{Dietrich2015} and \cite{Foucart2015} appears in good
agreement with the expected ``universal'' behaviour and its variance.

The $f_1$ frequencies also show a very good tight correlation with the
the tidal deformability. This was already reported in
Ref. \cite{Takami2015} (\cf Fig. 15 of \cite{Takami2015}) and we further
remark on this point in the right panel of Fig. \ref{Corr_f1}, where we
employ the same data of the left panel and fit it with a third-order
polynomial expansion of the dimensionless tidal deformability
$(\kappa^{^T}_2)^{1/5}$
\begin{equation}
\label{eq:empirical_f1_Lambda}
f_1 \approx c_0 + 
c_1~\left(\kappa^{^T}_2\right)^{\!\!1/5} +
c_2~\left(\kappa^{^T}_2\right)^{2/5} +
c_3~\left(\kappa^{^T}_2\right)^{3/5}\ {\mathrm{kHz}} \,,
\end{equation}
with the coefficients being given by
\begin{align}
&c_0 =  45.195 \,, &
&c_1 = -43.484 \,, &\nonumber\\
&c_2 =  14.653 \,, &
&c_3 = -1.6623 \,.
\label{eq:empirical_f1_coeffs_Lambda}
\end{align}
When considering only equal-mass binaries, the largest deviation from the
fit \eqref{eq:empirical_f1_Lambda} is $\simeq{7.1}\%$, while the average
is only $\simeq{2.3}\%$.

\begin{figure*}
\begin{center}
\includegraphics[width=0.85\columnwidth]
                {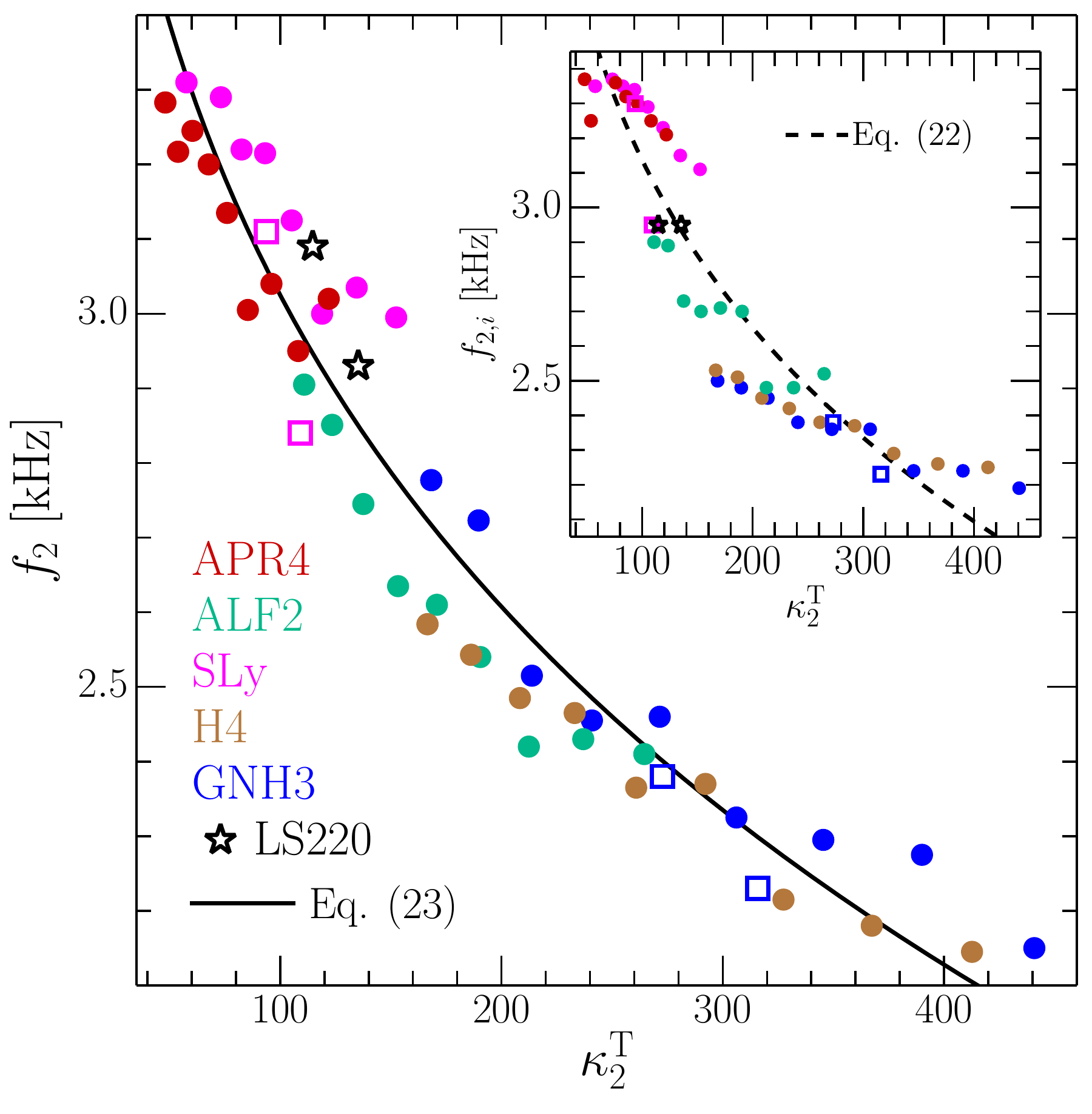}
\hskip 1.0cm
\includegraphics[width=0.85\columnwidth]
                {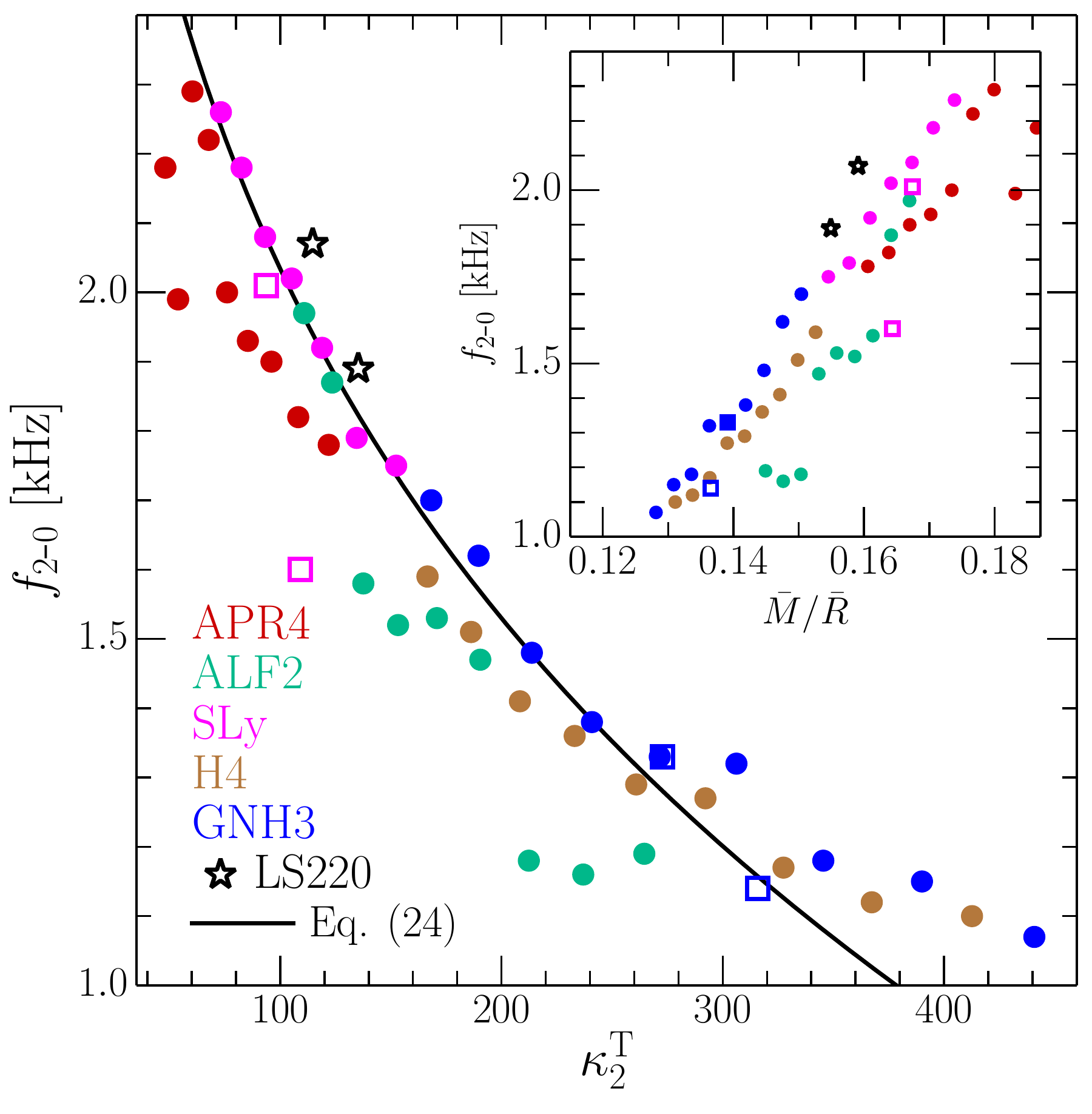}
\caption{\textit{Left panel:} Values of the $f_2$ peak frequencies shown
  as a function of the dimensionless tidal deformability $\kappa^{^T}_2$
  (we use the same symbol convention as in
  Fig. \ref{Corr_Mfmax}). Although there is a correlation, the scatter is
  much larger than for the $f_{\rm max}$ or $f_1$ frequencies (see
  Figs.~\ref{Corr_Mfmax} and \ref{Corr_f1}). The inset shows instead the
  $f_{2,i}$ frequencies for which the scatter is larger. \textit{Right
    panel:} Values of the $f_{2\mbox{-}0}$ peak frequencies shown as a
  function of the dimensionless tidal deformability $\kappa^{^T}_2$. In
  the inset, instead, the frequencies are reported as a function of the
  average compactness $\bar{M}/\bar{R}$, as done in
  Ref. \cite{Bauswein2015}.}
\label{Corr_f2}  
\end{center}
\end{figure*}

We next consider the correlations of the $f_{2,i}$ and $f_{2}$
frequencies with the stellar properties. We recall that the $f_{2,i}$ and
$f_{2}$ frequencies correspond to the same fundamental $\ell=2=m$ mode of
oscillation and the different denomination refers to whether the
frequencies are measured during the transient phase ($f_{2,i}$) or in the
quasi-periodic one ($f_{2}$). These frequencies are reported as a
function of the dimensionless tidal deformability in the left panel
Fig. \ref{Corr_f2}, where we report the data taken from all of our
simulations, including those that produce a black hole before 25 ms; a
similar figure was presented in Ref. \cite{Takami2015} (\cf Fig. 15 of
\cite{Takami2015}), but here we also provide a fitting function in terms
of the dimensionless tidal deformability, both for the $f_{2}$ and the
$f_{2,i}$ frequencies (see inset in the left panel of Fig. \ref{Corr_f2})
as
\begin{align}
\label{eq:newfit_f2i}
f_{2,i} &\approx 6.401 -1.299\,
\left(\kappa^{^T}_2\right)^{\!\!1/5}\quad \mathrm{kHz}
\,, &\\
\label{eq:newfit_f2}
f_{2} &\approx 5.832 -1.118\,
\left(\kappa^{^T}_2\right)^{\!\!1/5}\quad \mathrm{kHz}
\,. &
\end{align}
Relative to the equal-mass binaries, the largest (average) deviations
from the fit are $\simeq 11\%\ (4.9\%)$ for the $f_{2,i}$ frequencies and
$\simeq 9.0\%\ (3.0\%)$ for the $f_{2}$ frequencies.

In addition, we present in the right panel of Fig. \ref{Corr_f2} the
correlation between the $f_{2\mbox{-}0}$ frequencies and the tidal
deformability. We recall that the $f_{2\mbox{-}0}$ frequencies are simply
the difference between the $f_2$ frequencies, which are easy to measure
from the PSDs and the $f_{m=0}$ frequencies that can be estimated from
the analysis of the data of the simulations (\eg via the oscillation
frequencies of the central rest-mass density or of the lapse
function). In this sense, the $f_{2\mbox{-}0}$ frequencies are
straightforward to compute; yet, as discussed above, these do not always
correspond to a clearly visible peak in the total PSDs (\cf
Fig. \ref{PSD_5x7}). 
Using all of this data we obtain a linear fit in terms of shown as a
black solid line
\begin{align}
\label{eq:newfit_f20}
f_{2\mbox{-}0} &\approx 
5.424 -1.350\,\left(\kappa^{^T}_2\right)^{\!\!1/5}\quad
\mathrm{kHz}\,.
\end{align}
A similar correlation is exhibited by the compactness (see
\cite{Bauswein2015} and also the inset in the right panel of
Fig. \ref{Corr_f2}) and this is not particularly surprising given the
fundamental nature of the $f_{2\mbox{-}0}$ frequencies. Because the
correlation is weaker, the maximum and average deviations from the fit
are of $\simeq 36\%$ and $\simeq 8.3\%$, respectively.

We conclude this Section by reporting the fitting expressions for the
$f_{\rm spiral}$ frequencies as given by Eq.~(2) in \cite{Bauswein2015}
for equal-mass binaries with $\bar{M}=1.350\,\Msun$, 
\begin{equation}
\label{eq:fspir_135}
f_{\rm spiral} \approx 6.16 - 82.1~{\mathcal C} +
358~{\mathcal C}^2 \quad {\mathrm{kHz}}\,.
\end{equation}
No fitting expression was provided in \cite{Bauswein2015} for the
binaries with $\bar{M}=1.200, 1.500\,\Msun$, but it was not difficult to
reconstruct the behaviour of $f_{\rm spiral}$ through a quadratic
two-dimensional fit in terms of the compactness and average gravitational
mass of the binary, \ie 
\begin{align}
\label{eq:fspir_2D}
f_{\rm spiral} \approx & ~
d_0 + d_1~{\mathcal C} + d_2~{\mathcal C}^2 + & \nonumber\\
      &~ d_3~\bar{M} + d_4~\bar{M}^2 +
      d_5~{\mathcal C}~\bar{M} 
\quad \mathrm{kHz}\,, 
\end{align}
and where 
\begin{align}
&d_0 =  3.28 \,, &
&d_1 = -8.68 \,, &
&d_2 =  174  \,, & \nonumber\\
&d_3 = -2.34 \,, & 
&d_4 = 0.99  \,, &
&d_5 = -13.0 \,.  
\label{eq:2D_fit}
\end{align}
Expression \eqref{eq:fspir_2D} is the one used when indicating the
expected $f_{\rm spiral}$ frequencies in all the figures of this paper; a
comparison of the two-dimensional fit \eqref{eq:fspir_2D} with the
various fits suggested by Ref. \cite{Bauswein2015} will be presented in
Appendix \ref{appendix_c}.

\section{Conclusions}
\label{sec:conclusions}

The merger of a neutron-star binary is expected to be accompanied not
only by a highly energetic electromagnetic signal, but also by a strong
signal in GWs that will contain important signatures of the EOS of matter
at nuclear densities. These signatures are contained both in the inspiral
and in the post-merger signals and, following previous work of a number
of authors \cite{Baiotti:2010, Bernuzzi2012, Radice2013b, Radice2013c,
  Radice2015, Hotokezaka2016, Read2013, Bauswein2011, Stergioulas2011b,
  Bauswein2012, Hotokezaka2013c, Bauswein2014, Takami2014, Takami2015,
  Bernuzzi2015a, Palenzuela2015, Bauswein2015, Dietrich2015, Foucart2015}
we have here analysed the GW signal from merging neutron star binaries.

A particularly important goal of this work has been the analysis of the
post-merger signal, which presents a number of quite strong spectral
features that, just like spectral lines from the atmospheres of normal
stars, can be used to extract information on the physical properties of
the progenitor neutron stars. More specifically, we have tried to clarify
whether or not some of these spectral features can be correlated with the
properties of the stars before the merger in a way that is
quasi-universal, that is, only weakly dependent on the EOS.

To this scope we have simulated a large number of neutron-star binaries
in full general relativity and relativistic hydrodynamics. The binaries,
which are mostly with equal masses, have total masses as low as
$2.4\,M_{\odot}$ and as large as $3.0\,M_{\odot}$, spanning six different
nuclear-physics EOSs. Because of the close similarities in the EOSs, the
new set of binaries has been analysed together with the one already
presented in Refs. \cite{Takami2014, Takami2015}, thus providing a total
sample of 56 binaries, arguably the largest studied to date in general
relativity and with nuclear-physics EOSs.

A systematic analysis of the complete sample has allowed us to obtain a
rather robust picture of the spectral properties of the GW signal and
hopefully clarify a number of aspects that have been debated recently in
the literature. In essence, our most important findings can be summarised
as follows: 
\begin{itemize}
\item the instantaneous GW frequency when the amplitude reaches its first
  maximum is related quasi-universally with the tidal deformability of
  the two stars.

\item this correlation is observed for binaries with masses that do not
  differ of more than $20\%$.

\item the spectral properties vary during the post-merger phase with a
  marked difference between a transient phase lasting a few millisecond
  after the merger and a following quasi-stationary phase.

\item the most robust features of the post-merger signal pertain four
  frequencies: $f_1, f_{2}, f_3$, and $f_{2\mbox{-}0}$, where $f_{2}
  \simeq (f_1 + f_3)/2$ and $f_{2\mbox{-}0}$ is the result of a mode
  coupling.

\item when distinguishing the spectral peaks between these two phases, a
  number of ambiguities in the identification of the peaks disappear,
  leaving a rather simple and robust picture.

\item ``universal'' relations are found between the spectral features and
  the physical properties of the neutron stars.

\item for all of the correlations between the spectral features and the
  stellar properties, simple analytic expressions can be found either in
  terms of the dimensionless tidal deformability or of the stellar
  compactness.
\end{itemize}

When considered as a whole and in the light of recent direct detection of
GWs \cite{Abbott2016a}, these results open the exciting and realistic
prospects of constraining the EOS of nuclear matter via GW observations
of merging BNSs. At the same time, the robustness of these results needs
to be validated when different physical conditions are assumed for the
merging neutron stars. These include accounting a nonzero amount of
initial stellar spin \cite{Kastaun2013, Tsatsin2013, Bernuzzi2013,
  Tsokaros2015, Kastaun2014}, evaluating the modifications introduced by
ideal- and resistive-MHD effects \cite{Giacomazzo:2010,
  Dionysopoulou2015}, and assessing the impact that the $m=1$ shear
instability \cite{Ou06, Corvino:2010, Anderson2008, East2016} may have on
the post-merger spectrum when the instability is able to fully develop
and persist on secular timescales. Furthermore, while some examples of
unequal-mass binaries or of genuinely three-dimensional EOSs have been
considered here, a more systematic analysis needs to be performed; this
will be part of our future work.

\begin{acknowledgments}

We thank L. Baiotti, L. Bovard, M. Hanauske for useful comments and
discussions and D. Radice for help with the LS220 binaries. Partial
support comes from JSPS KAKENHI Grant Number 15H06813, from
``NewCompStar'', COST Action MP1304, from the LOEWE-Program in HIC for
FAIR, and the European Union's Horizon 2020 Research and Innovation
Programme under grant agreement No. 671698 (call FETHPC-1-2014, project
ExaHyPE). The simulations were performed on SuperMUC at LRZ-Munich and on
LOEWE at CSC-Frankfurt.
\end{acknowledgments}

\appendix

\section{Summary of stellar properties}
\label{appendix_a}

We report in Table \ref{tab:models} a summary of the main physical
properties of the binaries simulated in this work; some of the data has
already been provided in Ref. \cite{Takami2015}, but we report it also
here for completeness. The various columns denote the gravitational mass
ratio $q\equiv M_{_B}/M_{_A}$ at infinite separation, the average
gravitational mass $\bar{M}$ at infinite separation, the average radius
$\bar{R}$ at infinite separation, the ADM mass $M_\mathrm{ADM}$ of the
system at initial separation, the baryon mass $\bar{M}_\mathrm{b}$, the
compactness ${\mathcal C}$, the orbital frequency $f_\mathrm{orb}$ at the
initial separation, the total angular momentum $J$ at the initial
separation, the dimensionless moment of inertia $\bar{I}/\bar{M}^3$ at
infinite separation (which is tightly correlated with the compactness
\cite{Breu2016}), the $\ell=2$ dimensionless tidal Love number
$\bar{k}_2$ at infinite separation, the dimensionless tidal deformability
$\lambda/\bar{M}^5$, and the contact frequency $f_{\rm cont}$. All
quantities with a bar are defined as averages, \ie $\bar{\Psi} \equiv
(\Psi_{_A} + \Psi_{_B})/2$.

\begin{table*}
\begin{footnotesize}
\begin{tabular}{l|c|c|c|c|c|c|c|c|c|c|c|c|c|c}
\hline
\hline
model & EOS & $q$ & $\bar{M}$ & $\bar{R}$ 
& $M_\mathrm{ADM}$ & $\bar{M_\mathrm{b}}$
& $\bar{M}/\bar{R}$  
& $f_\mathrm{orb}$ & $J$ & $J/M_\mathrm{ADM}^2$ 
& $\bar{I}/\bar{M}^3$ & $\bar{k_2}$ & $\lambda/\bar{M}^5$ & $f_{\rm cont}$ 
\\
& & & $[\Msun]$ & $[\mathrm{km}]$ 
& $[\Msun]$ & $[\Msun]$
& & $[\mathrm{Hz}]$ & $[\Msun^2]$ 
&&&& $[\mathrm{Hz}]$ & \\
\hline
\mn{GNH3-q10-M1200} & GNH3 & $1.000$ & $1.200$ & $13.827$ & $2.3796$ & $1.2882$ & $0.12814$ & $268.89$ & $5.9899$ & $1.0579$ & $20.298$ & $0.12184$  & $2351.1$ & $1235.2$ \\
\mn{GNH3-q10-M1225} & GNH3 & $1.000$ & $1.225$ & $13.823$ & $2.4288$ & $1.3172$ & $0.13085$ & $2.7111$ & $6.1967$ & $1.0505$ & $19.575$ & $0.11970$  & $2080.2$ & $1248.6$ \\
\mn{GNH3-q10-M1250} & GNH3 & $1.000$ & $1.250$ & $13.817$ & $2.4780$ & $1.3464$ & $0.13358$ & $273.29$ & $6.4067$ & $1.0434$ & $18.890$ & $0.11753$  & $1842.4$ & $1262.1$ \\
\mn{GNH3-q10-M1275} & GNH3 & $1.000$ & $1.275$ & $13.810$ & $2.5271$ & $1.3756$ & $0.13632$ & $275.38$ & $6.6187$ & $1.0364$ & $18.237$ & $0.11531$  & $1633.0$ & $1275.7$ \\
\mn{GNH3-q10-M1300} & GNH3 & $1.000$ & $1.300$ & $13.801$ & $2.5763$ & $1.4050$ & $0.13908$ & $277.44$ & $6.8340$ & $1.0296$ & $17.614$ & $0.11305$  & $1448.1$ & $1289.4$ \\
\mn{GNH3-q10-M1325} & GNH3 & $1.000$ & $1.325$ & $13.790$ & $2.6255$ & $1.4345$ & $0.14187$ & $279.53$ & $7.0538$ & $1.0233$ & $17.019$ & $0.11075$  & $1284.7$ & $1303.2$ \\
\mn{GNH3-q10-M1350} & GNH3 & $1.000$ & $1.350$ & $13.777$ & $2.6746$ & $1.4641$ & $0.14468$ & $281.58$ & $7.2766$ & $1.0172$ & $16.450$ & $0.10841$  & $1139.9$ & $1317.3$ \\
\mn{GNH3-q10-M1375} & GNH3 & $1.000$ & $1.375$ & $13.762$ & $2.7237$ & $1.4938$ & $0.14753$ & $283.58$ & $7.5023$ & $1.0113$ & $15.904$ & $0.10602$  & $1011.5$ & $1331.7$ \\
\mn{GNH3-q10-M1400} & GNH3 & $1.000$ & $1.400$ & $13.744$ & $2.7729$ & $1.5236$ & $0.15040$ & $285.56$ & $7.7311$ & $1.0055$ & $15.381$ & $0.10359$  & $897.44$ & $1346.3$ \\
\mn{GNH3-q10-M1500} & GNH3 & $1.000$ & $1.500$ & $13.648$ & $2.9692$ & $1.6442$ & $0.16228$ & $293.15$ & $8.6778$ & $0.9843$ & $13.479$ & $0.09347$  & $553.71$ & $1408.3$ \\
\mn{GNH3-q09-M1300} & GNH3 & $0.926$ & $1.300$ & $13.797$ & $2.5763$ & $1.4052$ & $0.13912$ & $277.51$ & $6.8255$ & $1.0283$ & $17.607$ & $0.11297$  & $1445.0$ & $1289.9$ \\
\mn{GNH3-q08-M1275} & GNH3 & $0.821$ & $1.275$ & $13.789$ & $2.5274$ & $1.3770$ & $0.13653$ & $275.37$ & $6.5564$ & $1.0264$ & $18.205$ & $0.11483$  & $1613.6$ & $1278.6$ \\
\hline
\mn{H4-q10-M1200}   & H4   & $1.000$ & $1.200$ & $13.516$ & $2.3796$ & $1.2922$ & $0.13109$ & $268.83$ & $5.9885$ & $1.0576$ & $19.918$ & $0.12779$  & $2200.9$ & $1278.1$ \\
\mn{H4-q10-M1225}   & H4   & $1.000$ & $1.225$ & $13.525$ & $2.4288$ & $1.3213$ & $0.13373$ & $271.05$ & $6.1956$ & $1.0503$ & $19.247$ & $0.12571$  & $1959.3$ & $1290.1$ \\
\mn{H4-q10-M1250}   & H4   & $1.000$ & $1.250$ & $13.533$ & $2.4780$ & $1.3506$ & $0.13638$ & $273.25$ & $6.4058$ & $1.0432$ & $18.610$ & $0.12361$  & $1746.5$ & $1302.1$ \\
\mn{H4-q10-M1275}   & H4   & $1.000$ & $1.275$ & $13.539$ & $2.5271$ & $1.3799$ & $0.13904$ & $275.40$ & $6.6191$ & $1.0364$ & $18.004$ & $0.12147$  & $1558.3$ & $1314.1$ \\
\mn{H4-q10-M1300}   & H4   & $1.000$ & $1.300$ & $13.544$ & $2.5763$ & $1.4094$ & $0.14172$ & $277.52$ & $6.8356$ & $1.0299$ & $17.426$ & $0.11930$  & $1391.4$ & $1326.1$ \\
\mn{H4-q10-M1325}   & H4   & $1.000$ & $1.325$ & $13.548$ & $2.6255$ & $1.4390$ & $0.14440$ & $279.60$ & $7.0552$ & $1.0235$ & $16.873$ & $0.11708$  & $1243.1$ & $1338.3$ \\
\mn{H4-q10-M1350}   & H4   & $1.000$ & $1.350$ & $13.550$ & $2.6746$ & $1.4687$ & $0.14711$ & $281.61$ & $7.2770$ & $1.0173$ & $16.344$ & $0.11483$  & $1111.1$ & $1350.6$ \\
\mn{H4-q10-M1375}   & H4   & $1.000$ & $1.375$ & $13.550$ & $2.7237$ & $1.4985$ & $0.14983$ & $283.55$ & $7.5010$ & $1.0111$ & $15.837$ & $0.11253$  & $993.41$ & $1363.1$ \\
\mn{H4-q10-M1400}   & H4   & $1.000$ & $1.400$ & $13.548$ & $2.7728$ & $1.5284$ & $0.15258$ & $285.50$ & $7.7293$ & $1.0053$ & $15.350$ & $0.11019$  & $888.32$ & $1375.7$ \\
\mn{H4-q10-M1500}   & H4   & $1.000$ & $1.500$ & $13.518$ & $2.9692$ & $1.6494$ & $0.16384$ & $293.13$ & $8.6772$ & $0.9842$ & $13.580$ & $0.10046$  & $567.34$ & $1428.7$ \\
\hline
\mn{ALF2-q10-M1200} & ALF2 & $1.000$ & $1.200$ & $12.227$ & $2.3795$ & $1.3076$ & $0.14491$ & $268.76$ & $5.9850$ & $1.0570$ & $17.519$ & $0.13524$  & $1410.9$ & $1485.5$ \\
\mn{ALF2-q10-M1225} & ALF2 & $1.000$ & $1.225$ & $12.252$ & $2.4288$ & $1.3373$ & $0.14762$ & $271.00$ & $6.1923$ & $1.0497$ & $16.975$ & $0.13290$  & $1263.8$ & $1496.2$ \\
\mn{ALF2-q10-M1250} & ALF2 & $1.000$ & $1.250$ & $12.276$ & $2.4779$ & $1.3672$ & $0.15034$ & $273.16$ & $6.4014$ & $1.0425$ & $16.455$ & $0.13049$  & $1132.6$ & $1507.0$ \\
\mn{ALF2-q10-M1275} & ALF2 & $1.000$ & $1.275$ & $12.298$ & $2.5271$ & $1.3971$ & $0.15307$ & $275.12$ & $6.6103$ & $1.0351$ & $15.957$ & $0.12803$  & $1015.6$ & $1517.9$ \\
\mn{ALF2-q10-M1300} & ALF2 & $1.000$ & $1.300$ & $12.319$ & $2.5763$ & $1.4272$ & $0.15582$ & $277.26$ & $6.8274$ & $1.0287$ & $15.480$ & $0.12552$  & $911.00$ & $1529.0$ \\  
\mn{ALF2-q10-M1325} & ALF2 & $1.000$ & $1.325$ & $12.337$ & $2.6254$ & $1.4574$ & $0.15858$ & $279.36$ & $7.0475$ & $1.0224$ & $15.021$ & $0.12297$  & $817.44$ & $1540.1$ \\
\mn{ALF2-q10-M1350} & ALF2 & $1.000$ & $1.350$ & $12.353$ & $2.6746$ & $1.4877$ & $0.16136$ & $281.42$ & $7.2708$ & $1.0164$ & $14.581$ & $0.12037$  & $733.63$ & $1551.5$ \\
\mn{ALF2-q10-M1375} & ALF2 & $1.000$ & $1.375$ & $12.368$ & $2.7237$ & $1.5181$ & $0.16415$ & $283.45$ & $7.4970$ & $1.0106$ & $14.158$ & $0.11773$  & $658.50$ & $1563.0$ \\
\mn{ALF2-q10-M1400} & ALF2 & $1.000$ & $1.400$ & $12.380$ & $2.7728$ & $1.5487$ & $0.16697$ & $285.44$ & $7.7266$ & $1.0049$ & $13.750$ & $0.11505$  & $591.07$ & $1574.8$ \\
\mn{ALF2-q10-M1500} & ALF2 & $1.000$ & $1.500$ & $12.409$ & $2.9692$ & $1.6723$ & $0.17847$ & $293.14$ & $8.6758$ & $0.9841$ & $12.261$ & $0.10404$  & $383.01$ & $1624.3$ \\
\hline
\mn{SLy-q10-M1200}  & SLy  & $1.000$ & $1.200$ & $11.465$ & $2.3795$ & $1.3116$ & $0.15454$ & $268.59$ & $5.9804$ & $1.0562$ & $14.981$ & $0.10746$  & $812.79$ & $1636.0$ \\
\mn{SLy-q10-M1225}  & SLy  & $1.000$ & $1.225$ & $11.468$ & $2.4287$ & $1.3417$ & $0.15772$ & $2.7084$ & $6.1874$ & $1.0489$ & $14.478$ & $0.10507$  & $717.63$ & $1652.4$ \\
\mn{SLy-q10-M1250}  & SLy  & $1.000$ & $1.250$ & $11.469$ & $2.4779$ & $1.3720$ & $0.16092$ & $273.04$ & $6.3977$ & $1.0420$ & $14.000$ & $0.10266$  & $634.27$ & $1668.8$ \\
\mn{SLy-q10-M1275}  & SLy  & $1.000$ & $1.275$ & $11.470$ & $2.5271$ & $1.4024$ & $0.16413$ & $275.20$ & $6.6110$ & $1.0352$ & $13.545$ & $0.10025$  & $561.11$ & $1685.3$ \\
\mn{SLy-q10-M1300}  & SLy  & $1.000$ & $1.300$ & $11.469$ & $2.5763$ & $1.4330$ & $0.16736$ & $277.34$ & $6.8275$ & $1.0287$ & $13.113$ & $0.097835$ & $496.81$ & $1701.9$ \\
\mn{SLy-q10-M1325}  & SLy  & $1.000$ & $1.325$ & $11.468$ & $2.6254$ & $1.4637$ & $0.17060$ & $279.36$ & $7.0455$ & $1.0221$ & $12.702$ & $0.095415$ & $440.20$ & $1718.5$ \\
\mn{SLy-q10-M1350}  & SLy  & $1.000$ & $1.350$ & $11.465$ & $2.6745$ & $1.4946$ & $0.17386$ & $281.34$ & $7.2663$ & $1.0158$ & $12.309$ & $0.092993$ & $390.29$ & $1735.2$ \\
\mn{SLy-q10-M1375}  & SLy  & $1.000$ & $1.375$ & $11.461$ & $2.7237$ & $1.5256$ & $0.17714$ & $283.37$ & $7.4923$ & $1.0100$ & $11.935$ & $0.090570$ & $346.23$ & $1752.1$ \\
\mn{SLy-q10-M1400}  & SLy  & $1.000$ & $1.400$ & $11.456$ & $2.7728$ & $1.5568$ & $0.18043$ & $285.36$ & $7.7215$ & $1.0043$ & $11.577$ & $0.088149$ & $307.28$ & $1769.1$ \\
\mn{SLy-q10-M1500}  & SLy  & $1.000$ & $1.500$ & $11.424$ & $2.9692$ & $1.6832$ & $0.19387$ & $293.05$ & $8.6705$ & $0.9835$ & $10.292$ & $0.007850$ & $191.10$ & $1838.9$ \\
\mn{SLy-q09-M1300}  & SLy  & $0.926$ & $1.300$ & $11.467$ & $2.5763$ & $1.4333$ & $0.16739$ & $277.23$ & $6.8154$ & $1.0268$ & $13.115$ & $0.097828$ & $496.31$ & $1702.3$ \\
\mn{SLy-q08-M1275}  & SLy  & $0.821$ & $1.275$ & $11.457$ & $2.5273$ & $1.4043$ & $0.16432$ & $275.08$ & $6.5460$ & $1.0249$ & $13.560$ & $0.10018$  & $557.53$ & $1688.2$ \\
\hline
\mn{APR4-q10-M1200} & APR4 & $1.000$ & $1.200$ & $11.034$ & $2.3795$ & $1.3173$ & $0.16058$ & $268.60$ & $5.9801$ & $1.0562$ & $14.100$ & $0.10415$  & $650.23$ & $1732.9$ \\
\mn{APR4-q10-M1225} & APR4 & $1.000$ & $1.225$ & $11.043$ & $2.4287$ & $1.3478$ & $0.16379$ & $270.85$ & $6.1870$ & $1.0489$ & $13.651$ & $0.10197$  & $576.72$ & $1748.6$ \\
\mn{APR4-q10-M1250} & APR4 & $1.000$ & $1.250$ & $11.052$ & $2.4779$ & $1.3783$ & $0.16700$ & $273.05$ & $6.3973$ & $1.0419$ & $13.226$ & $0.099787$ & $512.14$ & $1764.3$ \\
\mn{APR4-q10-M1275} & APR4 & $1.000$ & $1.275$ & $11.060$ & $2.5271$ & $1.4090$ & $0.17022$ & $275.22$ & $6.6107$ & $1.0351$ & $12.821$ & $0.097595$ & $455.30$ & $1779.9$ \\
\mn{APR4-q10-M1300} & APR4 & $1.000$ & $1.300$ & $11.067$ & $2.5763$ & $1.4399$ & $0.17344$ & $277.24$ & $6.8245$ & $1.0282$ & $12.436$ & $0.095396$ & $405.19$ & $1795.5$ \\
\mn{APR4-q10-M1325} & APR4 & $1.000$ & $1.325$ & $11.073$ & $2.6254$ & $1.4709$ & $0.17667$ & $279.31$ & $7.0437$ & $1.0219$ & $12.070$ & $0.093194$ & $360.93$ & $1811.1$ \\
\mn{APR4-q10-M1350} & APR4 & $1.000$ & $1.350$ & $11.079$ & $2.6746$ & $1.5020$ & $0.17992$ & $281.37$ & $7.2665$ & $1.0158$ & $11.720$ & $0.090990$ & $321.78$ & $1826.7$ \\
\mn{APR4-q10-M1375} & APR4 & $1.000$ & $1.375$ & $11.084$ & $2.7237$ & $1.5334$ & $0.18317$ & $283.39$ & $7.4924$ & $1.0100$ & $11.387$ & $0.088786$ & $287.10$ & $1842.3$ \\
\mn{APR4-q10-M1400} & APR4 & $1.000$ & $1.400$ & $11.088$ & $2.7728$ & $1.5649$ & $0.18643$ & $285.39$ & $7.7218$ & $1.0043$ & $11.068$ & $0.086582$ & $256.33$ & $1858.0$ \\
\mn{APR4-q10-M1500} & APR4 & $1.000$ & $1.500$ & $11.096$ & $2.9692$ & $1.6925$ & $0.19960$ & $293.09$ & $8.6708$ & $0.9835$ & $9.9256$ & $0.077808$ & $163.73$ & $1921.1$ \\
\hline
\mn{LS220-q10-M1338}& LS220& $1.000$ & $1.338$ & $12.754$ & $2.6737$ & $1.4733$ & $0.15491$ & $2.8129$ & $7.2656$ & $1.0163$ & $14.489$ & $0.096525$ & $721.43$ & $1472.4$ \\
\mn{LS220-q10-M1372}& LS220& $1.000$ & $1.372$ & $12.732$ & $2.7414$ & $1.5150$ & $0.15910$ & $2.8404$ & $7.5772$ & $1.0082$ & $13.859$ & $0.093518$ & $611.61$ & $1494.8$ \\
\hline
\hline
\end{tabular}
\caption{All binaries evolved and their properties. The various columns
  denote the gravitational mass ratio $q\equiv M_{_B}/M_{_A}$ at infinite
  separation, the average gravitational mass $\bar{M}$ at infinite
  separation, the average radius $\bar{R}$ at infinite separation, the
  ADM mass $M_\mathrm{ADM}$ of the system at initial separation, the
  baryon mass $\bar{M}_\mathrm{b}$, the compactness ${\mathcal C}\equiv
  \bar{M}/\bar{R}$, the orbital frequency $f_\mathrm{orb}$ at the initial
  separation, the total angular momentum $J$ at the initial separation,
  the dimensionless moment of inertia $\bar{I}/\bar{M}^3$ at infinite
  separation, the $\ell=2$ dimensionless tidal Love number $\bar{k}_2$ at
  infinite separation, the dimensionless tidal deformability
  $\lambda/\bar{M}^5$ defined by $\lambda\equiv 2 \bar{k}_2 \bar{R}^5/3$
  and the contact frequency $f_{\rm cont}=\mathcal{C}^{3/2}/(2\pi
  \bar{M})$ \cite{Damour:2012}. The quantities with a bar are defined as
  averages, \ie $\bar{\Psi} \equiv (\Psi_{_A} + \Psi_{_B})/2$.
\label{tab:models}}
\end{footnotesize}
\end{table*}

Similarly, we report in Table \ref{tab:SNRs} the precise frequencies of
the main spectral properties of the GW signals computed here. In
particular, the various columns report the values of the frequency at
maximum amplitude $f_{\rm max}$, the low-frequency peak $f_{1}$, the
initial and stationary values of the largest peak frequencies $f_{2,i}$
and $f_2$, and the frequency of the quasi-radial axisymmetric ($m=0$)
mode $f_{m=0}$. For completeness, we recall that $f_{\rm max}$ is
measured from the evolution of the instantaneous GW frequency, $f_{1}$ is
estimated through the analytic expression \eqref{eq:empirical_f1} with
coefficients \eqref{eq:empirical_f1_coeffs}, $f_{2,i}$ is measured
through the spectrograms (\cf Fig. \ref{GWSpect_5x7}), $f_{2}$ is
measured through the full PSDs (\cf Fig. \ref{PSD_5x7}), while $f_{m=0}$
is measured from the evolution of the minimum value of the lapse function
$\alpha_{\rm min}$.

\section{Quasi-radial oscillations and the $f_{m=0}$ frequency}
\label{appendix_b}

\begin{figure*}
\begin{center}
\includegraphics[width=2.0\columnwidth]{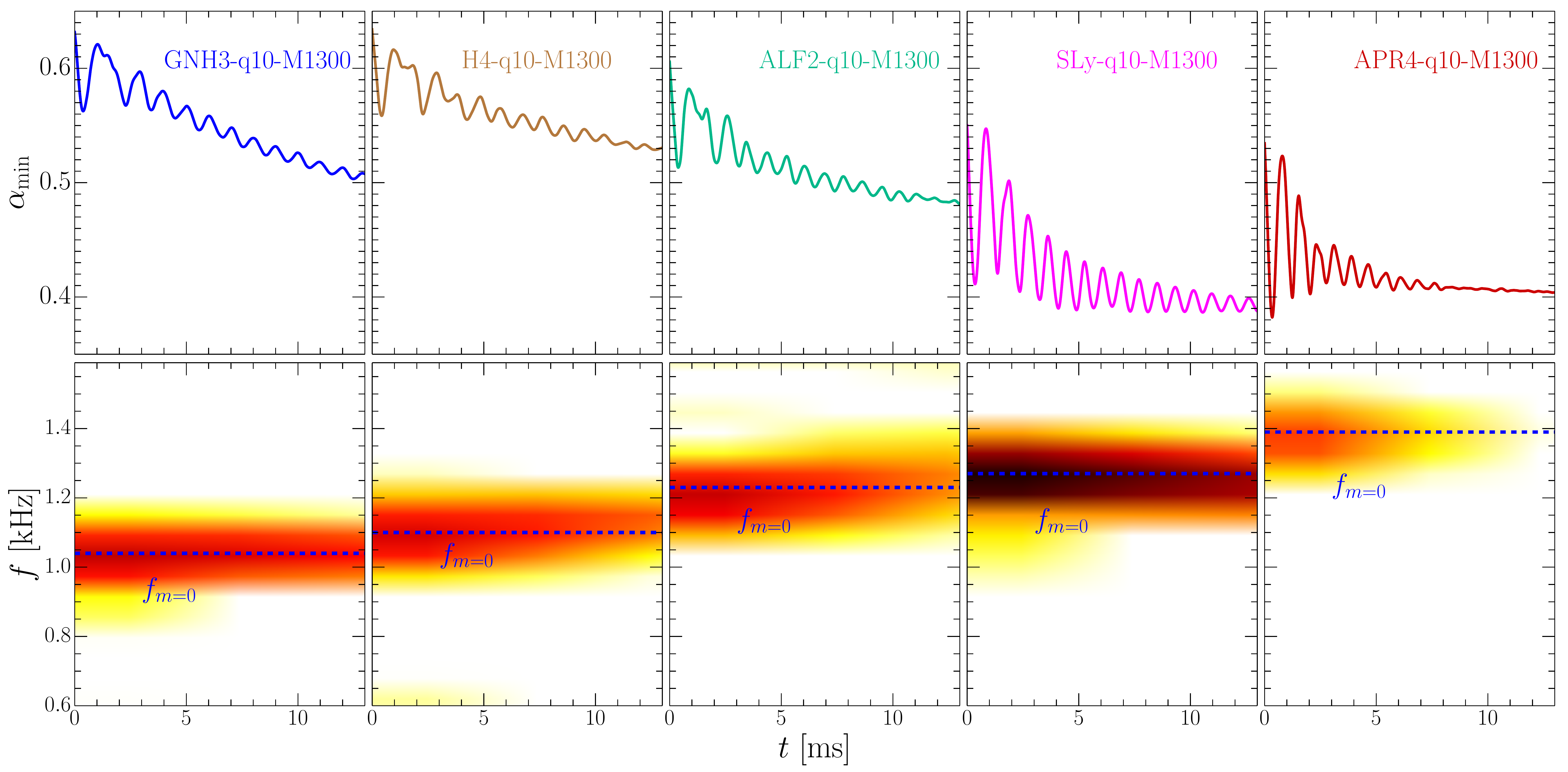}
\caption{Evolution of the minimum of the lapse function after the merger
  (top panels) and the corresponding spectrograms (bottom panels), for
  all the cold EOSs and binaries with mass $\bar{M}=1.300\Msun$. The
  horizontal blue dashed lines in the spectrograms indicate the
  fundamental quasi-radial oscillation frequencies $f_{m=0}$.}
\label{alp_1x5} 
\end{center}
\end{figure*}

As mentioned in Sect. \ref{sec:wp_ts}, the frequency $f_{2\mbox{-}0}$
refers to a coupling between the $\ell=2=m$ fundamental mode, which
yields the $f_2$ frequency, and a quasi-radial axisymmetric mode
$f_{m=0}$ \cite{Stergioulas2011b}. Hence, it can be computed as
$f_{2\mbox{-}0} \equiv f_2 - f_{m=0}$, once $f_{m=0}$ is measured.  There
are several different ways of doing this. A simple and convenient one is
to study the oscillations in the lapse function at the center of the
HMNS, which also corresponds to the minimum value of this function
$\alpha_{\rm min}(t)$, and to measure the oscillation period from a
spectrogram. This measure is very robust and a clear peak can be easily
isolated. As an example, we report in the top panels of
Fig. \ref{alp_1x5}, the evolution of $\alpha_{\rm min}(t)$ for the five
cold EOSs considered here and for the reference binary with mass
$\bar{M}=1.300\,M_{\odot}$. Also shown in Fig. \ref{alp_1x5}, but in the lower
panels, are the corresponding spectrograms. A rapid inspection of the
figure, both for $\alpha_{\rm min}(t)$ and for the frequencies, shows
that determining $f_{m=0}$ reliably is possible and straightforward. A
very similar behaviour is exhibited also by all the other binaries, which
we do not show for compactness.

\section{Two-dimensional fit of $f_{\rm spiral}$}
\label{appendix_c}

As mentioned in Sect. \ref{sec:wp_ts}, the frequency corresponding to
$f_{\rm spiral}$ cannot be measured reliably in our simulations without
the contamination of gauge effects. Hence, the values for these
frequencies can be only obtained from the analytic expression provided in
Ref. \cite{Bauswein2015}, which refers uniquely to binaries with
$\bar{M}=1.350\,\Msun$ [\cf Eq.~(2) of \cite{Bauswein2015}]. However, it
is not difficult to perform a two-dimensional fit of the data presented
in Fig. 7 of \cite{Bauswein2015} and obtain a fitting expression which is
given by our Eq. \eqref{eq:fspir_135} and can therefore be employed in
principle for any interpolating mass (we recall that the data in
\cite{Bauswein2015} refers to three sequences of binaries with
$\bar{M}=1.200, 1.350, 1.500\,\Msun$).

\begin{figure}
\begin{center}
\includegraphics[width=7.0cm]{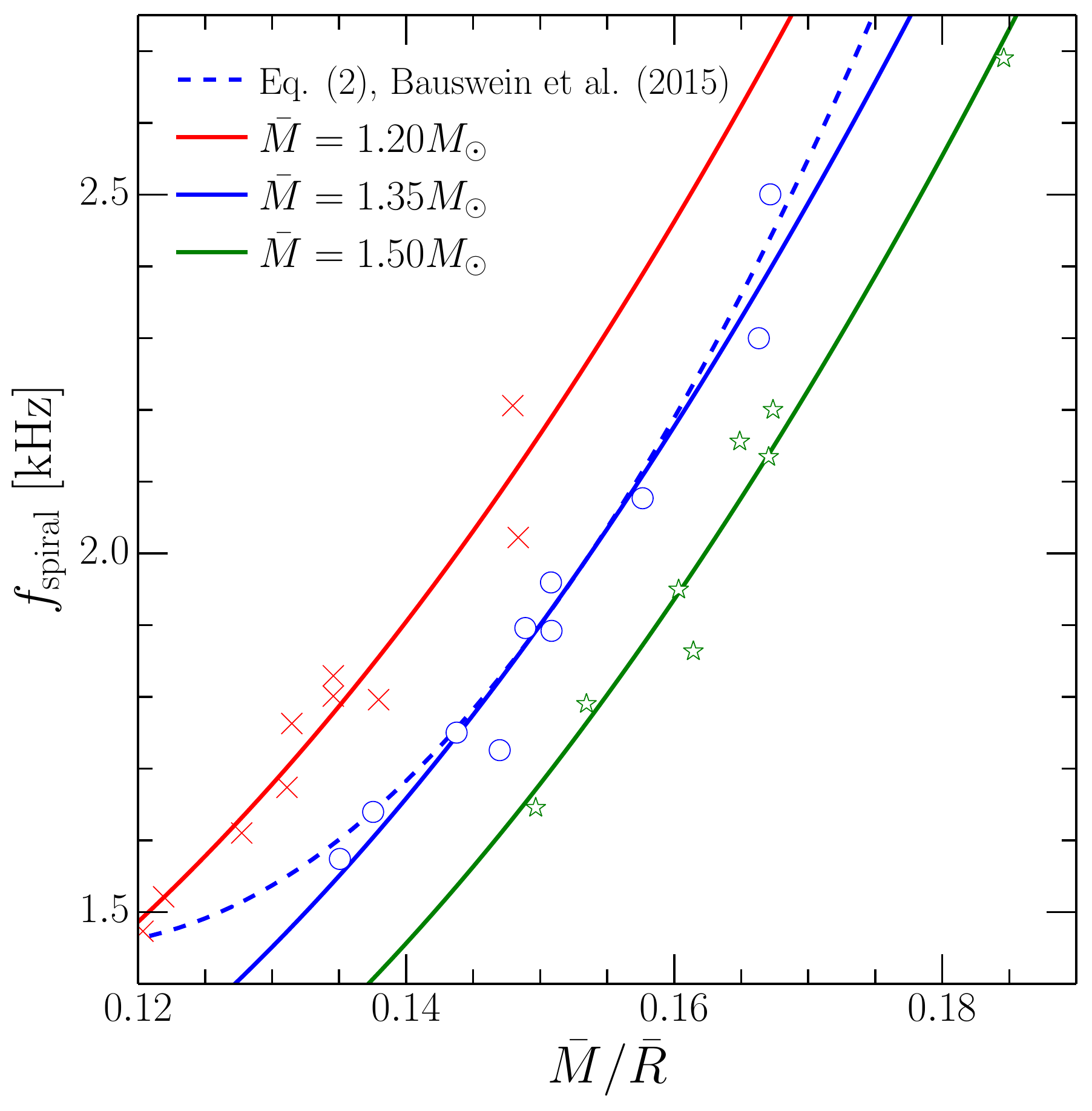}
\caption{Two-dimensional fit of the data presented in Fig. 7 of
  Ref. \cite{Bauswein2015} referring to the three sequences of binaries
  with $\bar{M}=1.200, 1.350, 1.500\,\Msun$, and shown with symbols of
  the different colour. The three solid lines represent our
  two-dimensional quadratic fit \eqref{eq:fspir_135} for the three
  sequences, while the blue dashed line is the one given by Eq. (2) of
  \cite{Bauswein2015}.}
\label{fig:2D_fit} 
\end{center}
\end{figure}

The goodness of our two-dimensional fit is shown in
Fig. \ref{fig:2D_fit}, where symbols of the different colour refer to the
data presented in Fig. 7 of Ref. \cite{Bauswein2015} and thus to to the
three sequences of binaries with mass $\bar{M}=1.200, 1.350,
1.500\,\Msun$, respectively. Shown instead with the three solid lines is
our fit for the three sequences, while the blue dashed line is the one
given by Eq. (2) of \cite{Bauswein2015}. Clearly, the two-dimensional fit
\eqref{eq:fspir_135} provides a very good representation of the data and
has therefore been used to estimate the values of $f_{\rm spiral}$ in the
various figures of this paper.

\begin{table}
\begin{footnotesize}
\begin{tabular}{l|c|c|c|c|c}
\hline 
\hline
model
 & $f_{\rm max}$ & $f_1$ & $f_{2,i}$ & $f_2$ & $f_{m=0}$ \\
 & $[\mathrm{kHz}]$ & $[\mathrm{kHz}]$ & $[\mathrm{kHz}]$
 & $[\mathrm{kHz}]$ & $[\mathrm{kHz}]$  \\
\hline
\mn{GNH3-q10-M1200} & $1.444$ & $1.45$ & $2.19$ & $2.15$ & $1.12$ \\
\mn{GNH3-q10-M1225} & $1.439$ & $1.57$ & $2.24$ & $2.28$ & $1.09$ \\
\mn{GNH3-q10-M1250} & $1.439$ & $1.60$ & $2.24$ & $2.30$ & $1.06$ \\
\mn{GNH3-q10-M1275} & $1.510$ & $1.65$ & $2.36$ & $2.33$ & $1.04$ \\
\mn{GNH3-q10-M1300} & $1.515$ & $1.65$ & $2.36$ & $2.46$ & $1.03$ \\
\mn{GNH3-q10-M1325} & $1.539$ & $1.68$ & $2.38$ & $2.46$ & $1.00$ \\
\mn{GNH3-q10-M1350} & $1.560$ & $1.70$ & $2.45$ & $2.52$ & $0.97$ \\
\mn{GNH3-q10-M1375} & $1.552$ & $1.80$ & $2.48$ & $2.72$ & $0.86$ \\
\mn{GNH3-q10-M1400} & $1.570$ & $1.90$ & $2.50$ & $2.78$ & $0.80$ \\
\mn{GNH3-q10-M1500} & $1.589$ & $--$   & $--$   & $--$   & $--$   \\
\mn{GNH3-q09-M1300} & $1.420$ & $1.50$ & $2.38$ & $2.38$ & $1.05$ \\
\mn{GNH3-q08-M1275} & $1.268$ & $1.20$ & $2.23$ & $2.23$ & $1.09$ \\
\hline
\mn{ALF2-q10-M1200} & $1.601$ & $1.73$ & $2.52$ & $2.41$ & $1.33$ \\
\mn{ALF2-q10-M1225} & $1.678$ & $1.81$ & $2.48$ & $2.43$ & $1.32$ \\
\mn{ALF2-q10-M1250} & $1.664$ & $1.82$ & $2.48$ & $2.42$ & $1.30$ \\
\mn{ALF2-q10-M1275} & $1.617$ & $1.86$ & $2.70$ & $2.54$ & $1.23$ \\
\mn{ALF2-q10-M1300} & $1.658$ & $1.88$ & $2.71$ & $2.61$ & $1.18$ \\
\mn{ALF2-q10-M1325} & $1.692$ & $1.95$ & $2.70$ & $2.64$ & $1.18$ \\
\mn{ALF2-q10-M1350} & $1.667$ & $1.99$ & $2.73$ & $2.75$ & $1.15$ \\
\mn{ALF2-q10-M1375} & $1.690$ & $2.00$ & $2.89$ & $2.85$ & $1.02$ \\
\mn{ALF2-q10-M1400} & $1.702$ & $2.02$ & $2.90$ & $2.91$ & $0.93$ \\
\mn{ALF2-q10-M1500} & $1.779$ & $--$   & $--$   & $--$   & $--$   \\
\hline
\mn{H4-q10-M1200}   & $1.441$ & $1.50$ & $2.25$ & $2.15$ & $1.15$ \\
\mn{H4-q10-M1225}   & $1.453$ & $1.60$ & $2.26$ & $2.18$ & $1.14$ \\
\mn{H4-q10-M1250}   & $1.473$ & $1.60$ & $2.29$ & $2.22$ & $1.12$ \\
\mn{H4-q10-M1275}   & $1.464$ & $1.64$ & $2.37$ & $2.37$ & $1.10$ \\
\mn{H4-q10-M1300}   & $1.489$ & $1.70$ & $2.38$ & $2.37$ & $1.09$ \\
\mn{H4-q10-M1325}   & $1.494$ & $1.70$ & $2.42$ & $2.47$ & $1.06$ \\
\mn{H4-q10-M1350}   & $1.529$ & $1.75$ & $2.45$ & $2.49$ & $1.04$ \\
\mn{H4-q10-M1375}   & $1.525$ & $1.85$ & $2.51$ & $2.54$ & $1.00$ \\
\mn{H4-q10-M1400}   & $1.537$ & $1.89$ & $2.53$ & $2.58$ & $0.94$ \\
\mn{H4-q10-M1500}   & $1.616$ & $--$   & $--$   & $--$   & $--$   \\
\hline
\mn{SLy-q10-M1200}  & $1.948$ & $1.84$ & $3.11$ & $3.00$ & $1.36$ \\
\mn{SLy-q10-M1225}  & $1.918$ & $1.90$ & $3.15$ & $3.04$ & $1.36$ \\
\mn{SLy-q10-M1250}  & $1.947$ & $2.00$ & $3.23$ & $3.00$ & $1.31$ \\
\mn{SLy-q10-M1275}  & $1.971$ & $2.05$ & $3.29$ & $3.13$ & $1.27$ \\
\mn{SLy-q10-M1300}  & $1.954$ & $2.13$ & $3.34$ & $3.22$ & $1.26$ \\
\mn{SLy-q10-M1325}  & $1.946$ & $2.25$ & $3.35$ & $3.22$ & $1.17$ \\
\mn{SLy-q10-M1350}  & $1.991$ & $2.37$ & $3.37$ & $3.29$ & $1.11$ \\
\mn{SLy-q10-M1375}  & $1.985$ & $2.65$ & $3.60$ & $3.58$ & $--$   \\
\mn{SLy-q10-M1400}  & $2.002$ & $--$   & $3.35$ & $3.31$ & $--$   \\
\mn{SLy-q10-M1500}  & $2.160$ & $--$   & $--$   & $--$   & $--$   \\
\mn{SLy-q09-M1300}  & $1.940$ & $2.20$ & $3.30$ & $3.11$ & $1.29$ \\
\mn{SLy-q08-M1275}  & $1.783$ & $1.90$ & $2.95$ & $2.84$ & $1.35$ \\
\hline
\mn{APR4-q10-M1200} & $1.959$ & $1.90$ & $3.21$ & $3.02$ & $1.43$ \\
\mn{APR4-q10-M1225} & $1.982$ & $1.95$ & $3.25$ & $2.95$ & $1.43$ \\
\mn{APR4-q10-M1250} & $1.992$ & $2.00$ & $3.30$ & $3.04$ & $1.40$ \\
\mn{APR4-q10-M1275} & $2.001$ & $2.10$ & $3.32$ & $3.01$ & $1.39$ \\
\mn{APR4-q10-M1300} & $2.035$ & $2.17$ & $3.36$ & $3.14$ & $1.36$ \\
\mn{APR4-q10-M1325} & $2.071$ & $2.41$ & $3.56$ & $3.20$ & $1.34$ \\
\mn{APR4-q10-M1350} & $2.056$ & $2.52$ & $3.60$ & $3.25$ & $1.31$ \\
\mn{APR4-q10-M1375} & $2.096$ & $2.70$ & $3.25$ & $3.22$ & $1.26$ \\
\mn{APR4-q10-M1400} & $2.109$ & $2.85$ & $3.37$ & $3.28$ & $1.19$ \\
\mn{APR4-q10-M1500} & $2.204$ & $--$   & $--$   & $--$   & $--$   \\
\hline
\mn{LS220-q10-M1338} & $1.703$ & $1.92$ & $2.95$ & $2.93$ & $1.06$ \\
\mn{LS220-q10-M1372} & $1.699$ & $2.05$ & $2.95$ & $3.09$ & $0.88$ \\
\hline
\hline
\end{tabular}
\caption{Values of the main spectral frequencies of the GW signals
  computed here: $f_{\rm max}, f_{1}$, $f_{2,i}$, $f_2$ and $f_{m=0}$. In
  the case of high-mass binaries the post-merger phase is too short to
  obtain reliable measures of the peak frequencies, which are therefore
  not reported.}
\label{tab:SNRs}
\end{footnotesize}
\end{table}

\bibliographystyle{apsrev4-1-noeprint}
%


\end{document}